\documentclass[11pt]{article}
\pdfoutput = 1

\usepackage[utf8]{inputenc}
\usepackage{color,graphicx}
\usepackage{epsfig}
\usepackage{amsmath}
\usepackage{verbatim}
\usepackage{amssymb}
\usepackage{mathabx}
\usepackage{stmaryrd}
\usepackage{physics}
\usepackage{amsfonts}
\usepackage{cite}
\usepackage{array}
\usepackage{setspace}
\usepackage{float}
\usepackage{url}
\usepackage{mathtools}
\usepackage{bbold}
\usepackage{slashed}
\usepackage{tikz}
\usepackage{graphicx}
\usepackage{epstopdf}
\usepackage{subfigure}
\usepackage[labelfont=bf,font={sf}]{caption}
\usepackage{pgfplots}
\usepackage{mathrsfs}
\usepackage{fancybox}
\usepackage{eurosym}
\usepackage{tcolorbox}
\usepackage{tensor}
\allowdisplaybreaks
\usepackage{simpler-wick}
\makeatletter
\pgfkeys{simplerwick,wickcolor/.store in=\swick@color,wickcolor=black}
\def\swick@end#1#2{
  \swick@setfalse@#1
  \tikzexternaldisable
  \begin{tikzpicture}[remember picture, baseline=(swick-close#1.base)]
    \node[use as bounding box, inner sep=0pt, outer sep=0pt] (swick-close#1) {$\displaystyle #2$};
  \end{tikzpicture}
  \tikz[remember picture, overlay]
    \draw[\swick@color] ($(swick-open#1.north) + (0, 3pt)$)
          -- ($(swick-open#1.base) + (0, \swick@offset) + #1*(0, \swick@sep)$)
          -- ($(swick-close#1.base) + (0, \swick@offset) + #1*(0, \swick@sep)$)
          -- ($(swick-close#1.north) + (0, 3pt)$);
  \tikzexternalenable}
\makeatother

\usepackage[margin = 2.2cm]{geometry}
\usepackage[ragged]{footmisc}
\usepackage[bookmarks=true,bookmarksnumbered=false,
hyperindex=true,bookmarksopen=true,hyperfigures=true,
colorlinks=true,linkcolor=black,citecolor=black,urlcolor=black,breaklinks]{hyperref}
\definecolor{webgreen}{rgb}{0, 0.5, 0}
\definecolor{webblue}{rgb}{0, 0, 0.5}
\definecolor{webred}{rgb}{0.5, 0, 0}
\definecolor{darkgreen}{rgb}{0,0.5,0}
\definecolor{ourorange}{rgb}{1, 0.502, 0}

\newcommand{\average}[1]{\left\langle #1 \right\rangle}

\def\ben{\begin{equation}}
\def\een{\end{equation}}

\let\a=\alpha \let\b=\beta \let\g=\gamma \let\d=\delta 
   
\let\l=\lambda     \let\r=v
\let\s=\sigma \let\t=\tau  \let\c=\chi

    \let\L=\Lambda
   
\let\C=\Chi

\def\nn{\nonumber}

\def\ba{\begin{array}}
\def\ea{\end{array}}

\def\dalemb#1#2{{\vbox{\hrule height .#2pt
       \hbox{\vrule width.#2pt height#1pt \kern#1pt
               \vrule width.#2pt}
       \hrule height.#2pt}}}

\newcommand{\bea}{\begin{eqnarray}}
\newcommand{\eea}{\end{eqnarray}}

\def\C{{{\mathbb{C}}}}
\def\H{{{\mathbb{H}}}}

\def\diag{{\rm diag}}

\renewcommand{\d}{\mathrm{d}}
\renewcommand{\i}{\mathrm{i}}
\renewcommand{\S}{\textsf{S}_0}
\renewcommand{\C}{\textsf{C}_0}
\renewcommand{\H}{\textsf{H}_0}
\newcommand{\Ss}{\textsf{S}}
\newcommand{\E}{\textsf{E}}

\newcommand{\be}{\begin{equation}}
\newcommand{\ee}{\end{equation}}

\numberwithin{equation}{section}

\title{Interior dynamics and the matrix elements of evaporating black holes}

\begin{document}

\thispagestyle{empty}
\begin{center}
     {\LARGE \bf 
    
  Microstructure in matrix elements
  }
    
    \vspace{0.4in}
    {\bf Andreas Blommaert$^1$ and Mykhaylo Usatyuk$^{2,3}$}

    \vspace{0.4in}
{$^1$Stanford Institute for Theoretical Physics,\\ Stanford University, Stanford, CA 94305 \\
$^2$Center for Theoretical Physics and Department of Physics, Berkeley, CA, 94720\\
$^3$Kavli Institute for Theoretical Physics,\\
University of California
Santa Barbara, CA 93106}
    \vspace{0.1in}
    
    {\tt ablommae@stanford.edu, musatyuk@berkeley.edu}
\end{center}

\vspace{0.4in}

\begin{abstract}
\noindent We investigate the simple model of Pennington, Shenker, Stanford and Yang for modeling the density matrix of Hawking radiation, but further include dynamics for EOW branes behind the horizon. This allows interactions that scatter one interior state to another, and also allows EOW loops. At strong coupling, we find that EOW states are no longer random; the ensemble has collapsed, and coupling constants encode the microscopic matrix elements of Hawking radiation. This suggests strong interior dynamics are important for understanding evaporating black holes, without any ensemble average. In this concrete model the density matrix of the radiation deviates from the thermal state, small off-diagonal fluctuations encode equivalences between naively orthogonal states, and bound the entropy from above. For almost evaporated black holes the off-diagonal terms become as large as the diagonal ones, eventually giving a pure state. We also find the unique analytic formula for all Renyi entropies.
\end{abstract}

\pagebreak
\setcounter{page}{1}
\tableofcontents
\setcounter{footnote}{0}
\section{Introduction}

Hawking argued that black holes evaporate into a mixed state of radiation \cite{Hawking:1975vcx}, even those formed from pure states. However, AdS/CFT implies that all black holes act as normal, unitary quantum mechanical systems when viewed from the outside \cite{Almheiri:2020cfm}. Hawking's notion of black hole evaporation must therefore be incomplete; black holes formed from pure states cannot evaporate into mixed states within AdS/CFT.

There has been significant recent progress on reconciling black hole evaporation with the constraints of unitarity \cite{Penington:2019npb,Almheiri:2019psf}. Instead of computing the state of Hawking radiation and evaluating its von Neumann entropy, the entropy can be computed using the replica trick and the gravitational path integral \cite{Almheiri:2019qdq,Penington:2019kki}. Replica wormholes contribute to the gravitational path integral, and give a unitary Page curve \cite{Page:1993df}. 

However, in these recent developments, the state of the radiation appears to be the same state that Hawking computed. Naively it seems inconsistent to have both a mixed state and a unitary Page curve. The reason is that, naively, gravity is dual to an ensemble average of unitary quantum systems; not one quantum system \cite{Saad:2019lba,Saad:2019pqd,Almheiri:2019qdq,Penington:2019kki,Marolf:2020xie,Stanford:2020wkf,Blommaert:2019wfy,Blommaert:2020seb,Pollack:2020gfa,Afkhami-Jeddi:2020ezh,Maloney:2020nni,Belin:2020hea,Cotler:2020ugk,Anous:2020lka,Chen:2020tes,Liu:2020jsv,Marolf:2021kjc,Meruliya:2021utr,Giddings:2020yes,Stanford:2019vob,Okuyama:2019xbv,Belin:2020jxr,Verlinde:2021jwu,Collier:2021rsn,Saad:2021uzi}. This gravity/ensemble duality \cite{Bousso:2020kmy} occurs when considering simple models of gravity, like pure JT gravity. 

We believe that more realistic models of quantum gravity, like those typically imagined in AdS/CFT, require no ensemble averaging. The price for unitarity, is a less simple bulk description. See \cite{Eberhardt:2018ouy, Eberhardt:2020bgq, Eberhardt:2021jvj} for a concrete recent example that averaging over string theories is not required.

In those realistic models, the Page curve is unitary, \emph{and} the density matrix of the radiation follows the standard rules of quantum mechanics; therefore the state does \emph{not} remain mixed. We would like to understand the bulk phenomena that explain the deviations of the density matrix from being maximally mixed. To accomplish this, we study the gravity description of one member of the ensemble dual to JT gravity with non-dynamical EOW branes \cite{Penington:2019kki}, focusing on the description of density matrix elements.

\begin{figure}
    \centering
    \begin{equation}
        \text{Re}(\rho_{i j})\,\, \raisebox{-22mm}{\includegraphics[scale=.4]{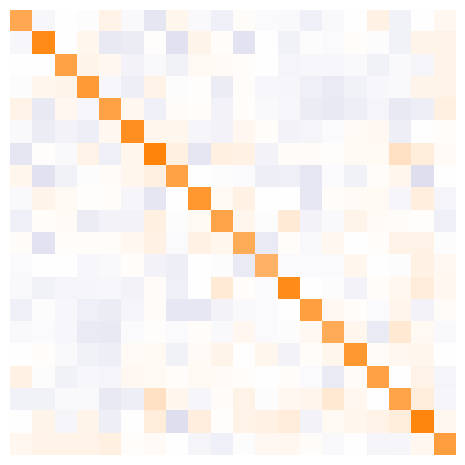}}\quad\quad\raisebox{-22mm}{ \includegraphics[scale=.4]{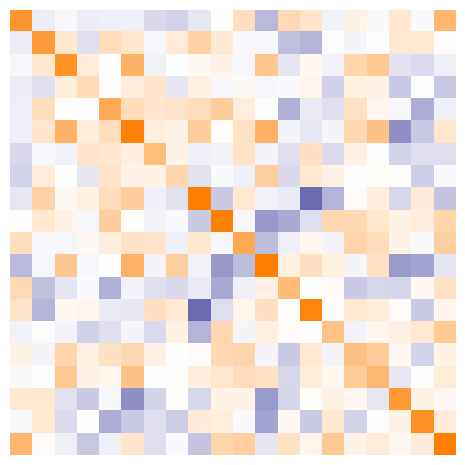}}\quad\quad\,\raisebox{-22mm}{\includegraphics[scale=.4]{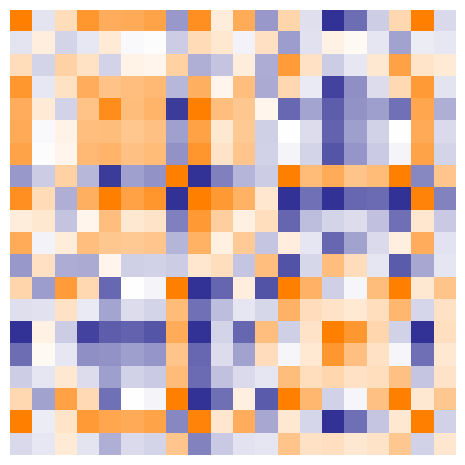}} \nn
    \end{equation}
    \caption{Normalized version of the density matrix \eqref{simpledensity} for $k=20$ and $e^{\Ss}=50\,,5\,,1$ (left, middle, right). Orange is positive and blue negative. Intensity of colors reflects the magnitude of individual matrix elements. Off-diagonal elements are less/not suppressed for black holes that have almost evaporated  (middle/right).}
    \label{fig:rhobis}
\end{figure}

The partition function for the ensemble dual to JT gravity with non-dynamical EOW branes, is \cite{Penington:2019kki}\footnote{See appendix D of \cite{Penington:2019kki}.}
\begin{equation}
    \mathcal{Z}=\int \d C\, \d C^\dagger\, \exp\bigg(-\Tr\Big(C^\dagger C\Big)\bigg)\int \d H\,\exp\bigg(-L\Tr\Big(V(H)\Big)\bigg)\,.\label{undef}
\end{equation}
One member of this ensemble is described by a $L\times L$ Hamiltonian $\H$, describing the bulk gravitational degrees of freedom; and furthermore by a $L\times k$ matrix $\C$, describing the interior states \cite{Penington:2019kki}. In section \ref{sect:discrete} we explain that in one member of the ensemble the density matrix of the radiation is essentially
\be
\rho=\sum_{i,j=1}^{k}\bra{\psi_j}\ket{\psi_i} \ket{i}\bra{j} = \sum_{i,j=1}^{k} \sum_{\alpha=1}^{e^{\Ss}} {\textsf{C}_{0}}^*_{\a j}\,{\textsf{C}_{0}}^{\,}_{\a i} \ket{i}\bra{j} = e^{\Ss}\sum_{i=1}^k\ket{i}\bra{i} +\mathcal{O}(e^{\Ss/2})\,.\label{simpledensity}
\ee
Here we took the microcanonical ensemble with $e^{\Ss}$ black hole states. We want to understand the gravity dual to a theory with fixed $\C$, which produces this density matrix; this is described in section \ref{sec:interior_dynamics}. The numerical structure of these matrix elements is discussed in section \ref{sect:discrete}. The state is plotted for fixed $\C$ in Fig. \ref{fig:rhobis}.\footnote{Similar plots were made for SYK with fixed couplings \cite{Stanford:2020wkf}, using a different representation for the density matrix \eqref{simpledensity}.}

The gravity interpretation of fixing the random Hamiltonians to $\H$ is a decoupled problem, aspects of which have been understood in \cite{drejorrit,Saad:2021rcu,Blommaert:2019wfy,Blommaert:2020seb}. This is not the focus of the present paper; nevertheless we include a short discussion on the associated extra ingredients in section \ref{sect:disc}.

\subsection{Summary, structure and main lessons}

\begin{figure}
    \centering
    \begin{equation}
    \bra{\psi_j}\ket{\psi_i}=\delta_{i j} \quad \raisebox{-10mm}{\includegraphics[width=14mm]{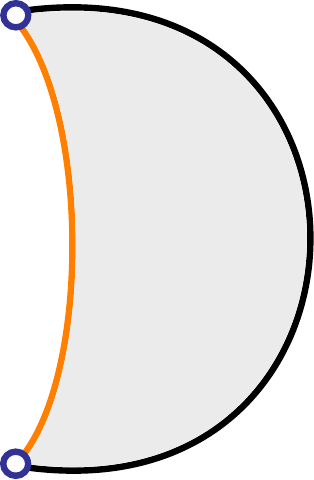}}\quad+\quad \raisebox{-10mm}{\includegraphics[width=14mm]{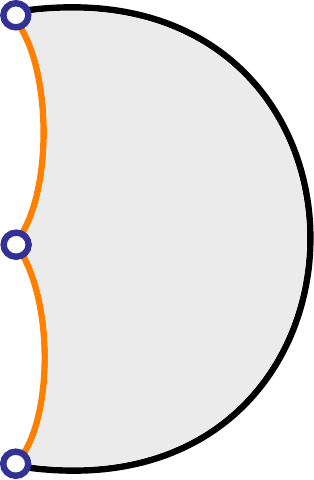}}\quad+\sum_{k_1=1}^k \quad \raisebox{-10mm}{\includegraphics[width=17.5mm]{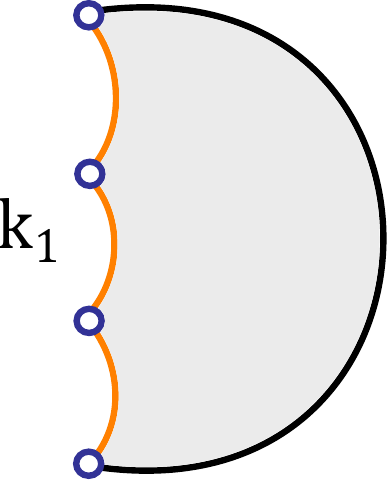}}\quad + \sum_{k_1,k_2=1}^k\quad \raisebox{-10mm}{\includegraphics[width=17.5mm]{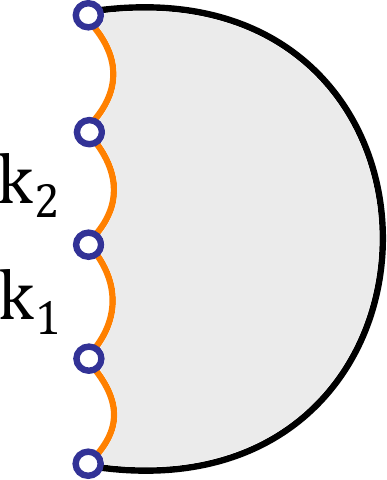}}\quad+\dots \nn
\end{equation}
    \caption{In the model of section \ref{sec:interior_dynamics} the off-diagonal terms come from EOW interactions where boundary particles can change flavor, the associated coupling constants depend on $\g$ and $\C$ and the summations are over different flavors of the intermediate boundary particles (external flavor labels $i$ and $j$ were left implicit).}
    \label{fig:corrections}
\end{figure}

In \textbf{section \ref{sec:interior_dynamics}} we are invited by gravity considerations to investigate deformations of the matrix integral \eqref{undef}\footnote{This is technically rather similar to deformations considered recently in \cite{drejorrit}. We briefly suppress the $H$ matrix integral.}
\be
\mathcal{Z} = \int d C d C^{\dagger} \exp\bigg(-\frac{1}{G}\Tr\Big(C^\dagger\, C\Big)- \g \Tr \Big(\C^{\dag}\,C + C^\dag\,\C \Big)\bigg)\,.\label{startstart}
\ee
The actual model for which we construct the gravitational dual in section \ref{sec:interior_dynamics} is slightly more complicated; here we simplify for presentation purposes. The gravitational interpretation of this deformation is to introduce scattering interactions from one EOW state into another, with coupling constants $g_{i j}$ that depend on $\g$ and $\C$. See \textbf{section \ref{sect:grav}} and Fig. \ref{fig:corrections} for gravity and \textbf{section \ref{sect:matint}} for the matrix integral.

Matrix elements are computed in this simplified description as ensemble averages of
\begin{equation}
    \bra{\psi_j}\ket{\psi_i}=\sum_{\a=1}^{e^\Ss}C^*_{\a j}C^{\,}_{\a i}\,.
\end{equation}

We are particularly interested in the models where the propagator $G$ takes the value $1/G=1+\g$. This interpolates between JT gravity with non-dynamical EOW branes \eqref{undef} for weak coupling $\g=0$
\begin{align}
    \mathcal{Z}=\int \d C\, \d C^\dagger \exp\bigg(-\Tr\Big(C^\dagger C\Big)\bigg)\,;\label{15}
\end{align}
and a gravity model with the matrix $C$ fixed to one member of the ensemble for strong coupling $\g=\infty$\footnote{Overall constants are irrelevant.}
\begin{align}
    \mathcal{Z}=\int \d C\, \d C^\dagger \exp\bigg(-\Tr\Big(C^\dagger C\Big)\bigg)\,\delta\Big(C-\C \Big)\delta\Big(C^\dagger-\C^\dagger \Big)\,.\label{16}
\end{align}
The stronger the interactions the less random the matrix $C$, and the more realistic the quantum gravity model under consideration. This is one key lesson of this work; in these two dimensional models, realistic gravity systems involve \textbf{strong interior dynamics}. See \textbf{section \ref{sect:collapse}}.

Furthermore, the \textbf{microscopic data} of the theory, here represented by the non-random matrix $\C$, are encoded in the specific \textbf{coupling constants} $g_{i j}$ for the EOW brane interactions. See \textbf{section \ref{sect:matint}}. 

For weak coupling $\g\ll1$, the matrix elements acquire small off-diagonal components
\begin{equation}
    \bra{\psi_j}\ket{\psi_i}=\frac{\delta_{i j}}{(1+\g)}+\frac{\g^2}{(1+\g)^2} \sum_{\alpha=1}^{e^{\Ss}} {\textsf{C}_{0}}^*_{\a j}\,{\textsf{C}_{0}}^{\,}_{\a i}\,.
\end{equation}
The first term, $\delta_{i j}$, corresponds to the usual rule for summing over EOW branes connected to asymptotic boundaries. If the brane flavor is the same on both ends of the boundary, the EOW particle can freely propagate. The second term accounts for scattering interaction between branes of different flavor, this leads to nonzero off-diagonal matrix elements \cite{Papadodimas:2012aq,Papadodimas:2013kwa,Penington:2019kki,Stanford:2020wkf}. See \textbf{section \ref{sect:grav}} and Fig. \ref{fig:corrections}.

In the strong coupling regime $\g\gg 1$, one recovers the non-random density matrix \eqref{simpledensity}. Although it may not be obvious, the off-diagonal terms are small relative to the diagonal ones in \eqref{simpledensity} when $e^{\Ss}\gg 1$. In realistic gravity models, without averaging, the density matrix \eqref{simpledensity} is \textbf{not maximally mixed}. The matrix $\C$ has dimensions $e^{\Ss}\times k$, which means that the rank of the density matrix is upper bounded by both $k$ and $e^{\Ss}$ \cite{Marolf:2020xie}; this suffices to understand the Page curve. See \textbf{section \ref{sect:31}} and \textbf{section \ref{sect:higher}}.

When we are ignorant about the microstructure of our system, we believe the states $\ket{\psi_i}$ are linearly independent; but in reality there are \textbf{equivalence relations} between them, there are null states. This is captured by numerical plots in Fig. \ref{fig:rhobis}. The leading order approximation to the density matrix retains only the dominant diagonal in Fig. \ref{fig:rhobis} (left) and suggests a maximally mixed state; however the smaller off-diagonal matrix elements become competitive when $k\,\,\propto\,\,e^{\Ss}$ and encode equivalence relations. When the black hole has almost evaporated  $e^{\Ss}\,\,\propto\,\,1$, the off-diagonal matrix elements are actually no longer suppressed, and the density matrix is far from being maximally mixed. Ultimately, for $e^{\Ss}=1$ the state becomes pure again. See Fig. \ref{fig:rhobis} (right) and \textbf{section \ref{sect:higher}}.

The Page transition is caused by the collective behavior of many small off-diagonal matrix elements of order $\mathcal{O}(e^{-\Ss/2})$ \cite{Papadodimas:2012aq,Papadodimas:2013kwa,Penington:2019kki,Stanford:2020wkf}, but the fact that the density matrix becomes pure again actually relies on the \textbf{off-diagonal matrix elements} ultimately \textbf{becoming large} at the end of evaporation.

In \textbf{section \ref{sect:entropy}} we obtain the unique analytic continuation for the Renyi entropies within the planar approximation, as an aside. In \textbf{section \ref{sect:disc}} we summarize our main findings, and discuss generalizations.

\section{Gravitational description} \label{sec:interior_dynamics}
In section \ref{sect:review} we review the simple model of \cite{Penington:2019kki} both from the gravity and matrix model perspective. In section \ref{sect:grav} brane interactions are introduced on the gravitational side, and it is shown that previously orthogonal inner products receive small overlaps due to these new interactions. In section \ref{sect:matint} the matrix model deformation that gives the brane interactions is introduced and analyzed, and in \ref{sect:collapse} the strong coupling limit is considered where the matrix $C$ is fixed to $\C$.\footnote{The Hamiltonians $H$ remains random throughout this section, see section \ref{sect:disc}.}

\subsection{Replica wormholes}\label{sect:review}
Following \cite{Penington:2019kki}, let $\ket{\psi_i}$ be the state of a black hole with a non-dynamical EOW brane with flavor $i$ behind the horizon. We would like to think of this state as modeling an interior mode of the radiation. Furthermore, let $\ket{i}$ be a basis of an auxiliary system, modeling the early outgoing Hawking modes.

Evaporating black holes can be emulated by considering the entangled, unnormalized state \cite{Penington:2019kki}
\begin{equation}
    \sum_{i=1}^k \ket{\psi_i}\otimes \ket{i}.
\end{equation}
Naively $\ket{\psi_i}$ form a $k$ dimensional basis; and concordantly this state is maximally mixed. Black hole evaporation is modeled by increasing $k$. Naively, increasing $k$ can make the entanglement between the black hole and radiation arbitrarily large, even exceeding the Bekenstein-Hawking entropy $\Ss$ - this is the maximal entanglement the black hole can have with the exterior. This contradiction is a version of the information paradox, similar to the Page curve.  

The resolution is that the states $\ket{\psi_i}$ are \emph{not} a $k$ dimensional basis; they have small nonzero overlaps that conspire to put an upper bound $e^{\Ss}$ on the dimension of their span \cite{Penington:2019kki,Marolf:2020rpm,Marolf:2021ghr,Stanford:2020wkf}. The way this gets diagnosed in \cite{Penington:2019kki}, is by computing the Renyi entropies associated with the reduced density matrix of the radiation\footnote{Throughout this section the density matrix is unnormalized.}
\begin{equation}
    \rho_{i j}=\braket{\psi_j}{\psi_i}\label{rij}.
\end{equation}

The state $\ket{\psi_i}$ is prepared by shooting in an EOW particle with flavor $i$ and mass $\mu$ at the boundary and then evolving for a thermal time $\beta/2$, giving the black hole some finite temperature. The bra works similarly, and because the EOW branes are non-dynamical there is no way for them to change flavors along their trajectory. To calculate inner products we apply the rules of the gravitational path integral. The inner product \eqref{rij} introduces boundary conditions consisting of an asymptotic boundary of length $\beta$ connected to two EOW branes of flavors $i$ and $j$. We sum over all geometries consistent with these boundary conditions. This results in the following gravitational amplitude, to leading order in $e^{\Ss}$

\begin{equation}
    \rho_{i j}=\delta_{i j}\quad \raisebox{-10mm}{\includegraphics[width=14mm]{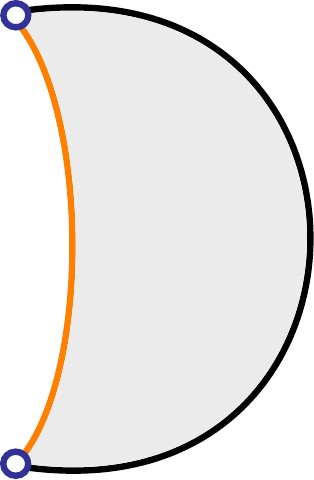}}\quad \propto\,\, \delta_{i j}\,e^{\Ss}.\label{rho}
\end{equation}
The black circle denotes an asymptotic JT gravity boundary of length $\beta$, and the orange line denotes the EOW particle with mass $\mu$.\footnote{We exclude brane labels on most gravitational diagrams for presentation purposes.} These amplitudes significantly simplify in the microcanonical ensemble, the black circle then denotes a microcanonical boundary condition \cite{Goel:2020yxl}. Details about gravitational amplitudes are gathered in appendix \ref{app:gravamp}.\footnote{We suppress subleading higher genus wormholes for reader comfort throughout.}

This looks like a maximally mixed state with the states $\ket{\psi_i}$ spanning a $k$ dimensional basis due to the $\delta_{i j}$. However, this conclusion changes when considering higher moments of matrix elements
\begin{equation}
    \rho_{i j}\,\rho_{k l}=\braket{\psi_j}{\psi_i}\braket{\psi_l}{\psi_k}.
\end{equation}
The particle with flavor $i$ can end up where one detects the outgoing particle with flavor $j$, however it could also end up where one detects the flavor $l$. Therefore we have two contributing Feynman diagrams
\begin{equation}
    \rho_{i j}\,\rho_{k l}=\delta_{i j}\,\delta_{k l}\quad \raisebox{-10mm}{\includegraphics[width=34mm]{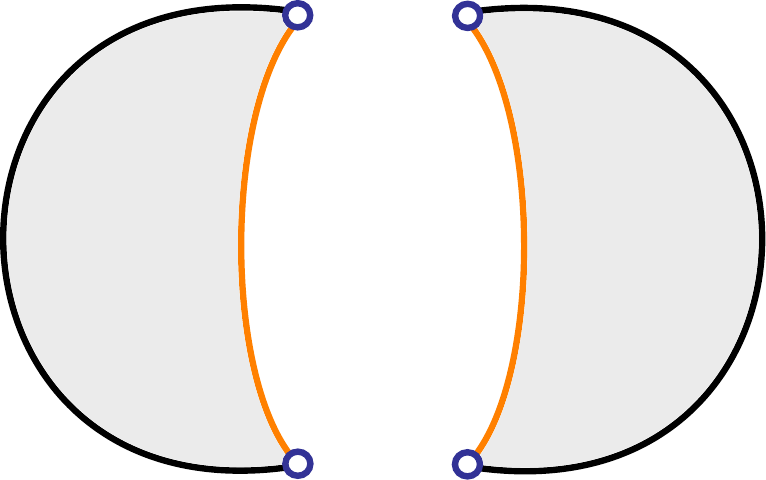}}\quad+\quad \delta_{i l}\,\delta_{k j}\quad \raisebox{-10mm}{\includegraphics[width=25.5mm]{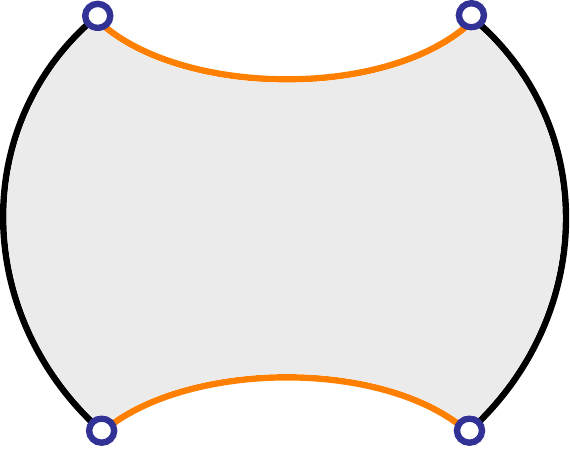}}\quad \propto\,\, \delta_{i j}\,\delta_{k l}\,e^{2\Ss}+\delta_{i l}\,\delta_{k j}\,e^{\Ss}.\label{rhorho}
\end{equation}
This is not simply the square of \eqref{rho}, because JT gravity with EOW branes is not one quantum system, but an ensemble of unitary quantum systems \eqref{undef}.

The expectation value for off-diagonal elements $\braket{\psi_j}{\psi_i}$ vanishes, but their standard deviation does not, as captured by the second geometry in \eqref{rhorho}. These smaller off-diagonal terms become important and severely constrain the span of $\ket{\psi_i}$ when $k$ exceeds $e^{\Ss}$. 

Consider for example the purity, which in the leading order approximation becomes \cite{Page:1993df}\footnote{There are corrections in the denominators from contracting with the normalization of the density matrix \cite{Stanford:2021bhl}.}$^,$\footnote{We define $R_n=\Tr(\rho^n)$ throughout, with $\rho$ normalized. These are not quite the Renyi entropies but play the same role.}
\begin{equation}
    R_2=\frac{1}{k}+\frac{1}{e^{\Ss}}\,.\label{pur}
\end{equation}
The second term stems from the replica wormhole in \eqref{rhorho} and dominates when $k\gg e^{\Ss}$, this effectively places an upper bound $e^{\Ss}$ on the dimension of the span of interior states $\ket{\psi_i}$. The ensemble averaged theory knows about the finite dimensionality of the interior state space.

The ensemble in question consists of a random $k\times L$ matrix $C$, describing the EOW brane degrees of freedom; and some random $L\times L$ Hamiltonian $H$ describing the bulk gravitational degrees of freedom \eqref{undef}. In this ensemble language, the random interior states and corresponding matrix elements are \cite{Penington:2019kki}\footnote{The $\pm$ means we multiply both signed gamma functions. We implicitly use the double scaled Hamiltonian throughout.}
\begin{align}
    \left|\psi_{i}\right\rangle&=\sum_{a, b=1}^{L} \Big(e^{-\beta H/2}\,\Gamma\left(\mu-1 / 2 \pm \mathrm{i} H^{1 / 2}\right)\Big)_{a b}\, C_{b i}^{\,}\,\ket{a}\nn\\
    \rho_{i j}&=\Big(C^\dagger\,\Gamma\left(\mu-1/2\pm \i H^{1/2}\right)^{1/2}e^{-\beta H}\,\Gamma\left(\mu-1/2\pm \i H^{1/2}\right)^{1/2}\,C\Big)_{j i}=\sum_{a,b=1}^L  C^*_{a j}\,\Big(\dots \Big)_{a b}\,C^{\,}_{b i}\,.\label{matint}
\end{align}
The vectors $\ket{a}$ represent a fixed rigid basis, analogous to the spin basis in SYK \cite{Kourkoulou:2017zaj}, the Hamiltonian $H$ in this basis is a matrix of random numbers. The brane states $\ket{\psi_i}$ are some random linear combination of the fixed basis states, specified by the matrix $C$. Wick contractions of $C$ are EOW branes in gravity,
see section \ref{sect:matint}.

\subsection{Interior dynamics}\label{sect:grav}
In a unitary gravity theory the density matrix elements are just numbers without any variance. This raises the question of what the gravity interpretation of these numbers is. We partially address this, by describing a gravity model where the ensemble over random matrices $C$ collapses to a fixed matrix $\C$.

The simple model of JT gravity with EOW branes, has mass $\mu$ boundary particles representing the branes, with action and boundary conditions \cite{Penington:2019kki}
\begin{equation}
    S=\mu \int \d s\,,\quad \partial_{n} \Phi=\mu\,, \quad K=0\,.\label{baction}
\end{equation}
The pieces of thermal boundary have fixed length boundary conditions \cite{Maldacena:2016upp,Engelsoy:2016xyb,Jensen:2016pah}, see also appendix \ref{app:gravamp}. We now enrich this model by allowing EOW interactions. EOW branes are boundaries on which two dimensional spacetimes end. This severely limits the set of EOW brane dynamics that we can introduce:
\begin{enumerate}
    \item There can be interaction vertices for $1\to 1$ EOW particle scattering where a particle of flavor $i$ scatters to a particle of flavor $j$, potentially accompanied by the emission or absorption of particles into the bulk spacetime. We restrict ourselves to one type of interaction, weighted with coupling constant $g_{i j}$
    \begin{equation}
        g_{i j}\quad \raisebox{-5.5mm}{\includegraphics[width=7mm]{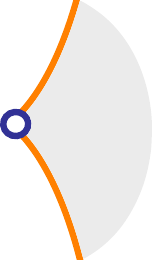}}\quad\text{or}\quad  g_{i j}\quad \raisebox{-5.5mm}{\includegraphics[width=7mm]{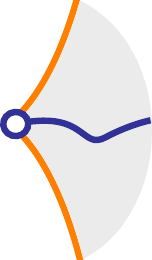}}\quad.
    \end{equation}
    Our specific choice for interaction vertices, which is of the first kind above, is detailed below and in appendix \ref{app:marking}.
    
    \item An EOW particle can propagate between two distinct points, these are either interaction vertices or points where the EOW particle ends on a bra or ket. We have the liberty to include an extra overall factor $G$ weighting every EOW propagator
    \begin{equation}
        G \, \raisebox{-9.5mm}{\includegraphics[width=8mm]{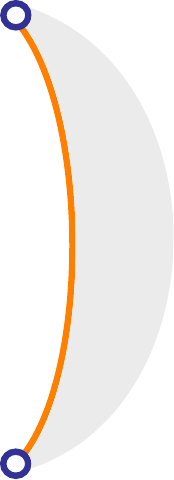}}\quad,
    \end{equation}
    which corresponds to adding a constant to the EOW brane action \eqref{baction}. This is somewhat ad-hoc, but entirely similar to the introduction of $S_\partial$ in \cite{Marolf:2020xie}. One can imagine more complicated quantum systems living on EOW branes which give these extra factors. We forsake the details since we are only interested in constructing a proxy for more general dynamical interiors.
    
    \item When $1\to 1$ interactions are included, there exist clearly also $0\to 2$ creation events, and $2\to 0$ annihilation events with the same coupling constants $g_{i j}$. Furthermore there is nothing preventing dynamical EOW branes from forming closed vacuum loops, making holes in the spacetimes. This exhausts all options for EOW brane interactions.\footnote{For example $1\to 2$ brane scattering interactions clearly cannot represent the boundary of some nonsingular spacetime.}
\end{enumerate}
Let us summarize the ingredients of our theory. There are $k$ flavors of EOW particles with the same mass $\mu$, and we have black hole states $\ket{\psi_i}$ for each particle flavor. There are interaction vertices where a particle of flavor $i$ turns into a particle with flavor $j$. Because this theory is dynamical, any number of interactions is allowed, and we must sum over all possible interactions when calculating amplitudes. Also, there are closed loops of EOW particles, with and without \cite{Gao:2021uro} interaction vertices on them.

The main new ingredient are interaction vertices where branes can change flavor; we must choose a specific way to model these interactions and deduce the corresponding JT gravity boundary conditions. For this it helps to think of JT gravity from the minimal string perspective \cite{brezin1993exactly,douglas1990strings,gross1990nonperturbative,Saad:2019lba,Mertens:2020hbs,Turiaci:2020fjj,Goel:2020yxl,douglasunpublished}.

Then each flavor of EOW particles corresponds with a D-brane, and interaction vertices correspond naturally with insertions of boundary operators $\mathcal{T}_{n\,i j}$; these are open string Tachyons stretching between D-branes, with Chan-Paton indices $i$ and $j$. Only the simplest of the chiral vertex operators $\mathcal{T}_{1\,i j}$, known as marking operators, have a known JT gravity interpretation \cite{Mertens:2020hbs,Mertens:2020pfe,Hosomichi:2008th,Kostov:2002uq} when stretching between D-branes with FZZT boundary conditions \cite{Saad:2019lba,Maldacena:2004sn,Fateev:2000ik,Ponsot:2001ng}.\footnote{Recently an educated guess was made for the interpretation of the other boundary operators \cite{Mertens:2020pfe}.} Marking operators then correspond with the $\beta=0$ limit of a fixed length boundary. This extends to mass $\mu$ boundary particle segments, relevant for EOW particles, since these are linear combinations of FZZT segments. See appendix \ref{app:marking}.

We have obtained sensible JT gravity boundary conditions for the EOW interaction vertices. This allows us to calculate any desired amplitude. We consider two examples to clarify the rules. 

\subsection*{Brane partition function}
We first consider something we call the D-brane partition function, consisting of all interacting EOW brane loops, and all spacetimes ending on them. These are the vacuum fluctuations of our system, they are modded out in all interesting calculations of matrix elements. Going through this first, lightens the presentation of matrix elements later.

The D-brane partition function $\mathcal{Z}$ has the following expansion\footnote{We suppress similar higher genus contributions with handles. The particle flavor indices are $p_1$, $p_2$ etcetera.}
\begin{equation} 
    \log \mathcal{Z}= \sum_{p_1=1}^k\,\, \raisebox{-5.5mm}{\includegraphics[width=16mm]{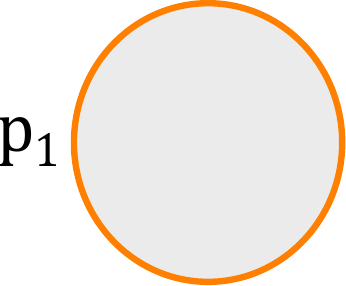}}\quad + \sum_{p_1=1}^k\,\, \raisebox{-8mm}{\includegraphics[width=16.5mm]{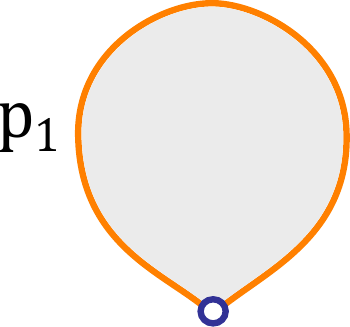}}\quad + \frac{1}{2} \sum_{p_1,p_2=1}^k\,\, \raisebox{-8mm}{\includegraphics[width=19.5mm]{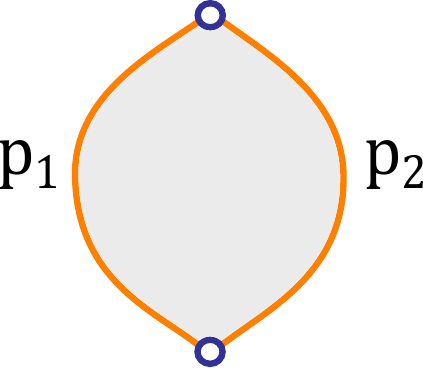}}\quad + \frac{1}{3} \sum_{p_1,p_2,p_3=1}^k\,\, \raisebox{-10mm}{\includegraphics[width=18mm]{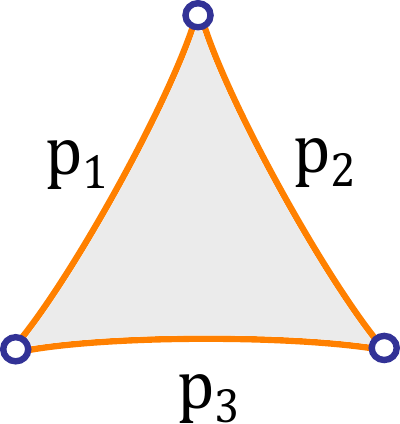}}\quad + \dots \label{dbranepart1}
\end{equation}
This is a sum over the number $n$ of scattering interactions, within each closed EOW particle loop. The $1/n$ symmetry factor is because cyclic permutations of the flavors describe the same Feynman diagram, which should be counted only once. The $\log$ reflects the fact that in the D-brane partition function $\mathcal{Z}$, we can have any number $m_n$ of those closed EOW particle loops with $n$ interactions and an identically ordered set of flavors; which are therefore indistinguishable. 

To compute the full D-brane partition function we must exponentiate the disks \eqref{dbranepart1} and then fill in bulk geometries; this includes cylinders connecting disks and more general wormhole topologies.

It is no accident that we use the same notation $\mathcal{Z}$ as for the matrix integral partition function, these are the same modulo disconnected spacetimes, like the sphere, which can be ignored; see section \ref{sect:matint}.

To compute these diagrams we must first isolate the EOW brane Feynman rules, meaning the factors $G$ and $g_{i j}$, from the basic JT gravity amplitudes; then simply compute the latter.

We find it convenient to denote gravitational boundary conditions by their matrix integral counterparts. The JT gravity calculation is insensitive to the flavors of the EOW particles, and only depends on the number of interactions $n$. In that case there are $n$ segments of mass $\mu$ boundary particles, separated by marked points; as explained in appendix \ref{app:marking} this corresponds with the operator insertion
\begin{equation}
    \Tr(\Gamma\left(\mu-1/2\pm \i H^{1/2}\right)^n)\,.
\end{equation}
The double scaled matrix integral dual of the loop without marking operators was deduced in \cite{Gao:2021uro} and reads\footnote{This looks remarkably similar up to signs to the relation of the FZZT loops with and without a marking operator.}
\begin{equation}
    \Tr \log (\Gamma\left(\mu+1/2\pm \i H^{1/2}\right))\,.
\end{equation}

When translating the partition function \eqref{dbranepart1} to gravity calculations, the EOW particle Feynman weights come out as prefactors for the gravity amplitudes. Because of the summation over flavor indices, the couplings $g_{i j}$ combine nicely into traces, and we obtain the gravitational ``boundary conditions''
\begin{equation}
    \log \mathcal{Z}=k\, G\, \Tr \log (\Gamma\left(\mu+1/2\pm \i H^{1/2}\right)) +\sum_{n=1}^\infty \frac{G^n}{n}\Tr\Big(g^n\Big) \Tr(\Gamma\left(\mu-1/2\pm \i H^{1/2}\right)^n)\,.\label{dbranepart2}
\end{equation}
There is an exponential of JT gravity boundaries in $\mathcal{Z}$, represented by the traces, we should sum over all spacetimes ending on them. The way to proceed with the gravity calculations is to use the general identity for expectation values of observables in any theory, crucial to understand D-branes \cite{Polchinski:1994fq,Saad:2019lba}
\begin{equation}
    \log \average{\exp(x)}=\sum_{m=1}^\infty \frac{1}{m!}\average{x^m}_\text{conn}\,.\label{poltrick}
\end{equation}
To good approximation one can then only include the exponential of disk shaped topologies, and annulus shaped topologies connecting to EOW particle boundaries \cite{Saad:2019lba}.\footnote{The others topologies have negative Euler character and therefore contribute negligibly assuming that $e^{\S}\gg 1$.} Because we will not need any detailed answers for the point we are trying to make in this paper, we omit the resulting expression.

\subsection*{Matrix elements}
Next consider matrix elements $\rho_{i j}=\bra{\psi_j}\ket{\psi_i}$, which give the gravitational boundary conditions discussed in \cite{Penington:2019kki}
\begin{equation}
    \bra{\psi_j}\ket{\psi_i}=\quad \raisebox{-10mm}{\includegraphics[width=14mm]{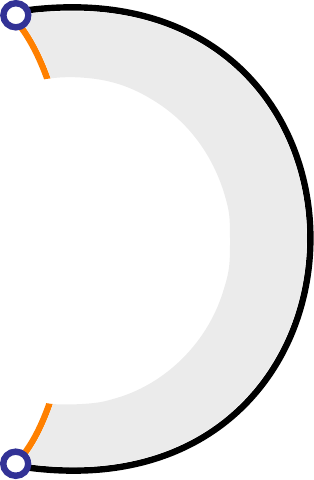}}\quad.
\end{equation}
As always we should sum over all possible Feynman diagrams ending on the boundary conditions. This includes EOW particle dynamics, and all gravitational spacetimes ending on the resulting diagrams.

Let us first consider the leading order amplitudes in small $g_{i j}$ perturbation theory, ignoring vacuum loops of EOW particles. For diagonal matrix elements one obtains up to order $g_{ij}$
\begin{align}
    \bra{\psi_i}\ket{\psi_i}&=\quad \raisebox{-10mm}{\includegraphics[width=14mm]{23_32.pdf}}\quad+\quad \raisebox{-10mm}{\includegraphics[width=14mm]{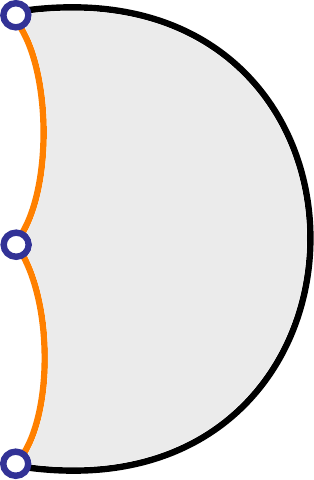}}\quad+\dots \nn\\&= G\, \Tr\Big(\Gamma\left(\mu-1/2\pm \i H^{1/2}\right)e^{-\beta H}\Big)+G^2\, g_{i i}\,\Tr\Big(\Gamma\left(\mu-1/2\pm \i H^{1/2}\right)^2e^{-\beta H}\Big)+\dots
\end{align}
where in the second equality we applied the EOW particle Feynman rules, and rewrote the gravity amplitudes by the corresponding observables in random matrix theory. For off-diagonal matrix elements, there is no leading contribution; however, crucially, there are contribution starting at linear order in $g_{i j}$
\begin{equation}
    \bra{\psi_j}\ket{\psi_i}=\quad \raisebox{-10mm}{\includegraphics[width=14mm]{220b_32.pdf}}\quad+\dots=G^2\, g_{i j}\,\Tr\Big(\Gamma\left(\mu-1/2\pm \i H^{1/2}\right)^2e^{-\beta H}\Big)+\dots
\end{equation}
So off-diagonal elements in the interacting theory are nonzero, unlike in the non-interacting model \eqref{rho}. This is the first sign that EOW particle interactions are important for understanding matrix elements in any non-random gravitational theory.

Higher orders in $g_{i j}$ are obvious; ignoring the vacuum loops this is an expansion in the number of scattering interactions on the EOW brane
\begin{equation}
    \bra{\psi_j}\ket{\psi_i}=\delta_{i j} \quad \raisebox{-10mm}{\includegraphics[width=14mm]{222a_32.pdf}}\quad+\quad \raisebox{-10mm}{\includegraphics[width=14mm]{222b_32.pdf}}\quad+\sum_{k_1=1}^k \quad \raisebox{-10mm}{\includegraphics[width=17.5mm]{222b_39.5.pdf}}\quad + \sum_{k_1,k_2=1}^k\quad \raisebox{-10mm}{\includegraphics[width=17.5mm]{222c_39.5.pdf}}\quad+\dots \label{probe}
\end{equation}
Note that unlike in \eqref{dbranepart1} there is no $1/n$ for the diagram with $n$ interactions, because the bra and the ket break the cyclic permutation symmetry. We translate this to pure gravity amplitudes by extracting the EOW particle Feynman weights. The JT gravity amplitudes are insensitive to the flavors, the sum over intermediate flavor indices combines the couplings into traces
\begin{equation}
    \bra{\psi_j}\ket{\psi_i}=\delta_{i j}\,G\,\Tr\Big(\Gamma\left(\mu-1/2\pm \i H^{1/2}\right)e^{-\beta H}\Big)+\sum_{n=1}^\infty G^{n+1}\,\Big(g^n\Big)_{i j} \Tr\Big(\Gamma\left(\mu-1/2\pm \i H^{1/2}\right)^{n+1}e^{-\beta H}\Big)\label{phiphiH}
\end{equation}

We should also include the effects of the EOW particle loops $\mathcal{Z}$. These are only truly vacuum loops if they are not connected to the probe boundaries \eqref{probe} via spacetime wormholes. In random matrix theory, denoting \eqref{phiphiH} by $\mathcal{O}$, matrix elements are really computed as $\average{\mathcal{O}\mathcal{Z}}/\average{\mathcal{Z}}$; and thus boundaries in \eqref{phiphiH} can connect, via spacetime wormholes, to boundaries in \eqref{dbranepart2}.

This means we will have contributions to \eqref{probe} with extra holes in the spacetimes; these holes are the EOW particle loops that the spacetime wormholes are connecting to, for example\footnote{The labels $p_1$, $p_2$ and $p_3$ on the sides of the inner triangle are suppressed for presentation purposes, and idem for $i$ and $j$.}
\begin{equation}
    \bra{\psi_j}\ket{\psi_i}\supset \frac{1}{3} \sum_{k_1,k_2=1}^k \sum_{p_1,p_2,p_3=1}^k \quad \raisebox{-10mm}{\includegraphics[width=21mm]{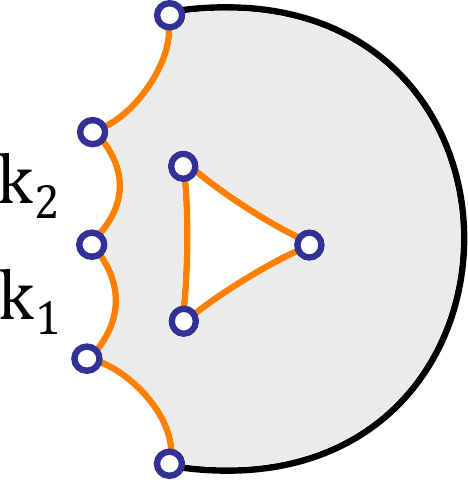}}\quad\,.\label{hole}
\end{equation}
There can be any number such holes in each portion of spacetime; if some of those have the same labels they are again indistinguishable. Notably one should not include the possibility of EOW particle loops that connect via spacetimes to each other, but not to any probe boundaries; those are normalized away with the $1/\average{\mathcal{Z}}$.

It is straightforward to compute the matrix elements order by order in small $g_{i j}$ perturbation theory; these are JT gravity amplitudes, which are known exactly. One all-encompassing example is discussed in appendix \ref{app:varia}.

The generalization to products of matrix elements like $\rho_{i j}\,\rho_{k l}=\braket{\psi_j}{\psi_i}\braket{\psi_l}{\psi_k}$, relevant for Renyi entropies, is clear. There are scattering interactions on all the EOW branes, and a nonzero answers for all values of $i,j,k$ and $l$. Furthermore, there can be holes due to EOW loops in all pieces of spacetime.

The Lorentzian interpretation is that there are interactions in the interior, where these EOW branes reside \cite{Kourkoulou:2017zaj,Penington:2019kki,Gao:2021uro}; schematically the associated Lorentzian spacetimes are
\begin{equation}
    \bra{\psi_j}\ket{\psi_i}\supset\quad \raisebox{-10mm}{\includegraphics[width=16.5mm]{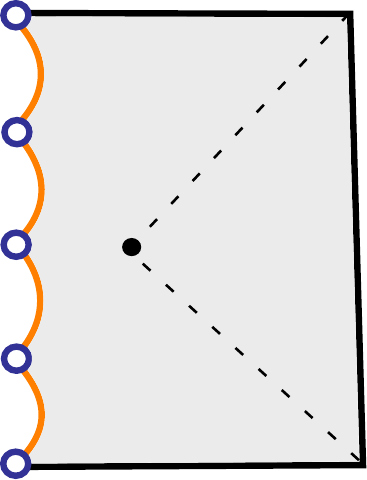}}\quad.
\end{equation}
Holes in the Euclidean spacetime \eqref{hole} are interpreted as associated with the spontaneous emission and absorption of \emph{open} baby universes, ofcourse there is also still the spontaneous emission and absorption of closed baby universes; associated with the spacetime wormholes which encode eigenvalue correlation \cite{Saad:2019pqd}.

\subsection{Dual matrix integral}\label{sect:matint}
We next discuss the matrix integral dual of this model of JT gravity with interaction EOW particles. Using this, we can consider large $g_{i j}$. This strong coupling limit selects one member $\C$ of the ensemble.

Consider first the matrix dual to JT gravity with non-dynamical EOW branes with partition function \eqref{undef}, and with matrix elements \eqref{matint}
\begin{align}
    \mathcal{Z}&=\int \d C\, \d C^\dagger\, \exp\bigg(-\Tr\Big(C^\dagger C\Big)\bigg)\int \d H\,\exp\bigg(-L\Tr\Big(V(H)\Big)\bigg) \label{nondyn1}\\
    \bra{\psi_j}\ket{\psi_i}&=\Big(C^\dagger\,\Gamma\left(\mu-1/2\pm \i H^{1/2}\right)^{1/2}e^{-\beta H}\,\Gamma\left(\mu-1/2\pm \i H^{1/2}\right)^{1/2}\,C\Big)_{j i}\,.\label{nondyn2}
\end{align}
The Gaussian integral over the complex matrix $C$ reduces to standard Wick contractions
\begin{equation}
    \wick { \c1 C_{a i} \c1 C^*_{b j}}=\delta_{i j}\delta_{a b}\,.\label{wick}
\end{equation}
For one matrix elements $\rho_{i j}=\bra{\psi_j}\ket{\psi_i}$, the ensemble average over $C$ therefore gives
\begin{equation}
    \bra{\psi_j}\ket{\psi_i}=\delta_{i j}\,\Tr(\Gamma\left(\mu-1/2\pm \i H^{1/2}\right)\,e^{-\beta H})\,.
\end{equation}
For two copies of the matrix element $\rho_{i j}\,\rho_{k l}=\braket{\psi_j}{\psi_i}\braket{\psi_l}{\psi_k}$, summing over Wick contractions gives\footnote{\begin{align}
    \bra{\psi_j}\ket{\psi_i}&=\sum_{a,b=1}^L \wick{\c2 C^*_{a j}\,\Big(\dots \Big)_{a b}\,\c2 C_{b i}}=\delta_{i j}\,\sum_{a=1}^L\Big(\dots \Big)_{a a}=\delta_{i j}\,\Tr\Big(\dots \Big)\\
    \bra{\psi_j}\ket{\psi_i}\braket{\psi_l}{\psi_k}&=\sum_{a,b,c,d=1}^L \wick{\c2 C^*_{a j}\,\Big(\dots \Big)_{a b}\,\c2 C_{b i}}\,\wick{\c2 C^*_{c l}\,\Big(\dots \Big)_{c d}\,\c2 C_{d k}}+\sum_{a,b,c,d=1}^L \wick{\c2 C^*_{a j}\,\Big(\dots \Big)_{a b}\,\c1 C_{b i}\,\c1 C^*_{c l}\,\Big(\dots \Big)_{c d}\,\c2 C_{d k}}=\dots\nn
\end{align}
}
\begin{align}
    \braket{\psi_j}{\psi_i}\braket{\psi_l}{\psi_k}&=\delta_{i j}\,\delta_{k l}\Tr(\Gamma\left(\mu-1/2\pm \i H^{1/2}\right)\,e^{-\beta H})\Tr(\Gamma\left(\mu-1/2\pm \i H^{1/2}\right)\,e^{-\beta H})\nn\\&\qquad\qquad+ \delta_{i l}\,\delta_{k j}\Tr(\Gamma\left(\mu-1/2\pm \i H^{1/2}\right)\,e^{-\beta H}\,\Gamma\left(\mu-1/2\pm \i H^{1/2}\right)\,e^{-\beta H})\,.
\end{align}
As explained in appendix \ref{app:gravamp}, these operator insertions in the $H$ matrix integral correspond in gravity with the EOW brane geometries shown in formula \eqref{rho} and \eqref{rhorho} respectively. Each $\Gamma\left(\mu-1/2\pm \i H^{1/2}\right)$ represents a geodesic boundary segment with a mass $\mu$ EOW particle, and each factor $e^{-\beta H}$ corresponds with a fixed length $\beta$ segment. Segments inside each trace form a closed loop. We see that every Wick contraction of matrix elements of $C$ becomes an EOW particle propagator in gravity. Therefore, \eqref{nondyn1} and \eqref{nondyn2} corresponds indeed with JT gravity with non-dynamical EOW branes, see appendix D in \cite{Penington:2019kki}.

We claim that the model of section \ref{sect:grav} corresponds with the deformed matrix integral
\begin{align}
    \mathcal{Z}&= \int \d C\, \d C^\dagger \exp(-\frac{1}{G}\Tr\Big(C^\dagger C\Big)+\g\Tr\Big(\C^\dagger\, \Gamma\left(\mu-1/2\pm \i H^{1/2}\right)^{1/2}\,C\Big)+\text{c.c.})\nn\\&\qquad\qquad\qquad\qquad\qquad\qquad\qquad\qquad\qquad\qquad\qquad\qquad\int \d H\,\exp\bigg(-L\Tr\Big(V(H)\Big)\bigg)\,,\label{model}
\end{align}
where the coupling constants for EOW particle scattering are determined by $g$ and $\C$
\begin{equation}
    g_{i j}=\g^2 \frac{1}{L}\Big(\C^\dagger\,\C\Big)_{j i}\,.\label{couplcoupl}
\end{equation}
For now the matrix $\C$ are just complex numbers parameterizing the coupling constants, but eventually it will represent the matrix that the $C$ ensemble collapses to, see section \ref{sect:collapse}. We next prove that \eqref{model} is equivalent to the gravity model of section \ref{sect:grav}, by computing the same quantities and matching them. 

\subsection*{Brane partition function}
We start with the brane partition function $\mathcal{Z}$. The integral over $C$ in \eqref{model} remains a simple Gaussian, this can immediately be computed by completing the square\footnote{We discard some irrelevant overall normalization constant which depends only on $G$.}
\begin{equation}
    \mathcal{Z}=\int \d H\,\exp\bigg(-L\Tr\Big(V(H)\Big)\bigg)\exp\bigg(G\,\g ^2\Tr(\C\,\C^\dagger\,\Gamma\left(\mu-1/2\pm \i H^{1/2}\right))\bigg)\label{first}\,.
\end{equation}
Note that the exponential in the integrand is not $U(L)$ invariant because of the presence of the matrices $\C\,, \C^\dagger$. Concordantly, the gravitational interpretation of this insertion is not immediately clear; the dictionary between operator insertions in random matrix theory and gravitational boundary conditions concerns only $U(L)$ invariant operators. To transform the above into $U(L)$ invariant operator insertions we can diagonalize the random matrix $H$ with random unitaries $U$ \cite{mehta2004random}\footnote{The following manipulations follow those previously used in \cite{drejorrit}, where more details can be found.}
\begin{equation}
    H=U\,\Lambda\,U^\dagger\,,\quad \Lambda=\diag(\lambda_1,\dots,\lambda_L)\,,\quad \d H=\d U\, \d \l_1\dots d\l_L \prod_{\alpha<\beta}^L(\l_\alpha-\l_\beta)^2\,,\label{diag}
\end{equation}
with $\d U$ the Haar measure on $U(L)$; and then explicitly compute the integral over the random unitaries. We are led in \eqref{first} to calculate the following integral over Haar random unitaries
\begin{equation}
    \int \d U\, \exp\bigg(G\,\g^2\Tr\Big(\C\,\C^\dagger\,U\,F(\L)\,U^\dagger\Big)\bigg)\,,\quad F(\Lambda)=\Gamma\left(\mu-1/2\pm \i \L^{1/2}\right)\,.\label{hc1}
\end{equation}

This Harish-Chandra integral can be computed exactly \cite{HarishChandra,ItzyksonZuber}. However, to link with the expansion of section \ref{sect:grav}, it is actually more practical to instead apply \eqref{poltrick} to correlators in the Haar ensemble \cite{drejorrit}
\begin{equation}
    \log \average{\exp\bigg(G\,\g^2\Tr\Big(\C\,\C^\dagger\,U\,F(\L)\,U^\dagger\Big)\bigg)}=\sum_{n=1}^\infty \frac{G^n}{n!}\,\g^{2n}\bigg\langle\Tr\Big(\C\,\C^\dagger\,U\,F(\L)\,U^\dagger\Big)^n\bigg\rangle_\text{conn}\,.\label{randu}
\end{equation}
The unitary group integrals on the right evaluate to a double sum over permutations\footnote{\begin{equation}
    \Tr_\s\Big(A^{n}\Big)=\prod_{\s_{i}} \operatorname{Tr}\left(A^{\ell\left(\s_{i}\right)}\right)\,,\quad \ell(\s_i)\text{ length cycles of }\s
\end{equation}}
\be \label{Weingarten}
\int \d U\,\Tr\Big(\C\,\C^\dagger\,U\,F(\L)\,U^\dagger\Big)^n = \sum_{\s,\t \,\in\, \mathcal{S}_n} \Tr_\s\Big(\Big(\C\,\C^\dagger\Big)^n\Big) \Tr_{\t} \Big(F(H)^n\Big)\, {\rm Wg}(\s/\t,L)\,.
\ee
This sum over permutations is weighted with Weingarten functions ${\rm Wg}(\sigma,L)$, which are known explicitly \cite{gu2013moments,Roberts:2016hpo}. We are interested in continuum JT gravity, where $L=\infty$. Using the leading large $L$ behavior of Weingarten functions; one checks that these correlators of Haar unitaries reduce, for large $L$, to the Wick contractions $\s=\t$ generated by a Gaussian complex matrix $U$ with variance $L$ \cite{Blommaert:2020seb}. Explicitly
\begin{equation}
\int \d U\,\Tr\Big(\C\,\C^\dagger\,U\,F(\L)\,U^\dagger\Big)^n \overset{\text{large $L$}}{=} \frac{1}{L^{n}} \sum_{\sigma\,\in\, \mathcal{S}_n} \Tr_\s\Big(\Big(\C\,\C^\dagger\Big)^n\Big) \Tr_{\s} \Big(F(H)^n\Big)\,.
\end{equation}

\begin{figure}[t]
    \centering
    \includegraphics[width=134mm]{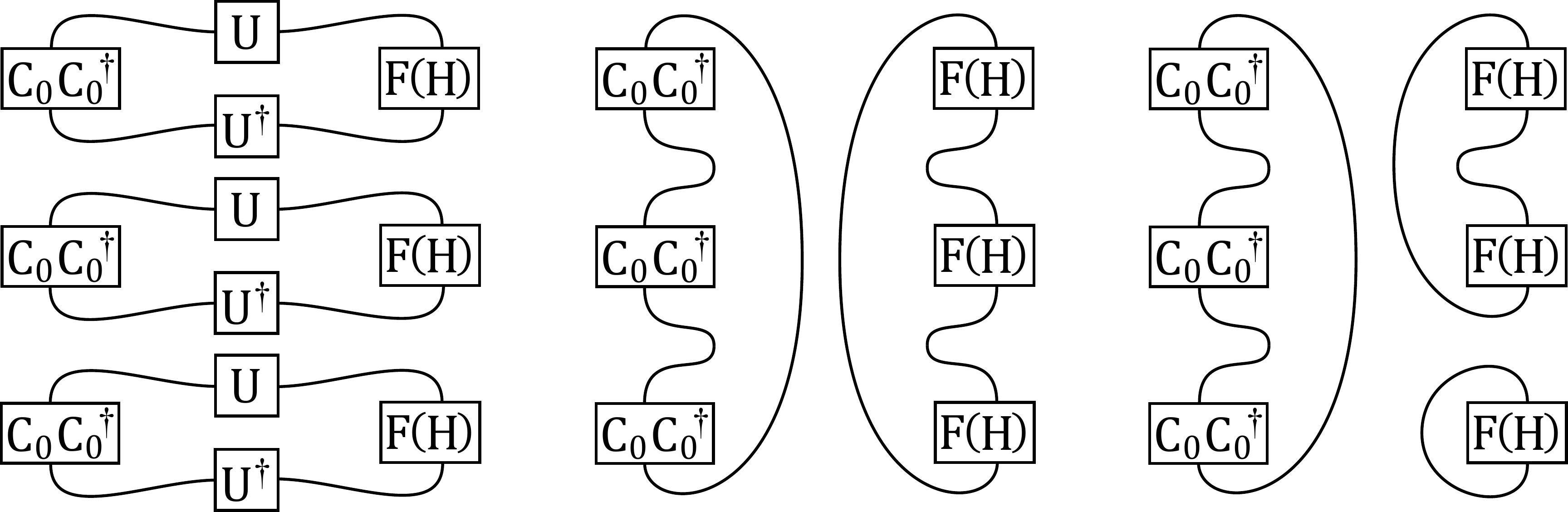}
    \caption{Haar random unitaries. Wires contract indices, integrating over Haar random unitaries corresponds with inserting complete sets of wire states, Weingarten functions weight each bra-ket combination. Dominant terms for large $L$ have identical bra and ket (middle), the subleading terms have different bra and ket (right).}
    \label{fig:weingarten}
\end{figure}

In this Gaussian approximation one therefore finds 
\begin{equation}
    \bigg\langle\Tr\Big(\C\,\C^\dagger\,U\,F(\L)\,U^\dagger\Big)^n\bigg\rangle_\text{conn}=(n-1)!\,\frac{1}{L^n}\Tr\Big(\Big(\C\,\C^\dagger\Big)^n\Big) \Tr(\Gamma\left(\mu-1/2\pm \i H^{1/2}\right)^n)\,.\label{aproxxx}
\end{equation}
The combinatorial prefactor counts the cycles of length $n$ in $\mathcal{S}_n$, only these contribute to the connected correlator. Combining this with \eqref{randu} and using the dictionary between the couplings \eqref{couplcoupl}, the brane partition function \eqref{first} becomes
\begin{equation}
    \mathcal{Z}=\int \d H\,\exp\bigg(-L\Tr\Big(V(H)\Big)\bigg)\exp\bigg(\sum_{n=1}^\infty \frac{G^n}{n}\Tr\Big(g^n\Big) \Tr(\Gamma\left(\mu-1/2\pm \i H^{1/2}\right)^n)\bigg)\,.\label{iden1}
\end{equation}
This matches the D-brane partition function of our JT gravity theory with interacting EOW particles \eqref{dbranepart2}, modulo the first term in \eqref{dbranepart2}, which represents EOW loops without interactions. Those have nothing to do with the $C$ matrix integral, and are therefore of little interest here. We can include them by deforming the potential $V(H)$ by the first term in \eqref{dbranepart2}, before doing the $C$ or $U$ integrals in \eqref{model}.

In summary, we have shown that the deformed matrix integral partition function \eqref{model} is equivalent for weak coupling $\g$, with the D-brane partition function of JT gravity with interacting EOW branes discussed in section \ref{sect:grav}. At strong coupling there are modification to this picture, as the approximation \eqref{aproxxx} breaks down; we discuss in the discussion section \ref{sect:disc} how this affect the gravitational description. We define the strongly coupled version of the gravity theory in section \ref{sect:grav} as the gravity dual to \eqref{model}.

\subsection*{Matrix elements}

We now check that the matrix model \eqref{model} also reproduces the matrix elements \eqref{phiphiH} that we computed in the gravitational theory, hence establishing the full fledged duality of the matrix and gravity models.

These matrix elements are computed in the deformed matrix integral \eqref{model} via the dictionary \eqref{nondyn2}
\begin{align}
    \bra{\psi_j}\ket{\psi_i}=\,&\frac{1}{\mathcal{Z}}\int \d C\, \d C^\dagger \exp(-\frac{1}{G}\Tr\Big(C^\dagger C\Big)+\g\Tr\Big(\C^\dagger\, \Gamma\left(\mu-1/2\pm \i H^{1/2}\right)^{1/2}\,C\Big)+\text{c.c.})\\&\int \d H\,\exp\bigg(-L\Tr\Big(V(H)\Big)\bigg)\,\Big(C^\dagger\,\Gamma\left(\mu-1/2\pm \i H^{1/2}\right)^{1/2}e^{-\beta H}\,\Gamma\left(\mu-1/2\pm \i H^{1/2}\right)^{1/2}\,C\Big)_{j i}\nn\,.
\end{align}
We first calculate the Gaussian integral over the matrix $C$. The two matrices $C$ and $C^\dagger$ in the operator insertion can either contract with one another, or with the $C$ and $C^\dagger$ in the deformation term in \eqref{model}. The first Wick contraction gives
\begin{align}
    \bra{\psi_j}\ket{\psi_i}\supset\,&\frac{1}{\mathcal{Z}}\int \d H\,\exp\bigg(-L\Tr\Big(V(H)\Big)\bigg)\,\exp\bigg(G\,\g^2\Tr(\C\,\C^\dagger\,\Gamma\left(\mu-1/2\pm \i H^{1/2}\right))\bigg)\nn\\&\qquad\qquad\quad\qquad\qquad\qquad\qquad\qquad\qquad\qquad \delta_{i j}\,G\,\Tr\Big(\Gamma\left(\mu-1/2\pm \i H^{1/2}\right)e^{-\beta H}\Big)\,,\label{cont1}
\end{align}
where $\mathcal{Z}$ represents \eqref{first}. The Wick contraction with the exponential deformation gives
\begin{align}
    \bra{\psi_j}\ket{\psi_i}\supset\,&\frac{1}{\mathcal{Z}}\int \d H\,\exp\bigg(-L\Tr\Big(V(H)\Big)\bigg)\,\exp\bigg(G\,\g^2\Tr(\C\,\C^\dagger\,\Gamma\left(\mu-1/2\pm \i H^{1/2}\right))\bigg)\nn\\&\qquad\qquad\qquad\qquad G^2\,\g^2\Big(\C^\dagger\,\Gamma\left(\mu-1/2\pm \i H^{1/2}\right)e^{-\beta H}\Gamma\left(\mu-1/2\pm \i H^{1/2}\right)\,\C\Big)_{j i}\,.\label{cont2}
\end{align}

Neither of these integrands is $U(L)$ invariant, to give this integral a gravitational interpretation we again diagonalize the Hamiltonian as in \eqref{diag}, and compute the resulting integral over random unitaries $U$ via the Gaussian large $L$ approximation as explained around \eqref{aproxxx}. For the first contribution \eqref{cont1} this is the same calculation as \eqref{hc1} because the operator insertion on the second line of \eqref{cont1} \emph{is} $U(L)$ invariant, therefore we find
\begin{align}
    \bra{\psi_j}\ket{\psi_i}\supset\,&\frac{1}{\mathcal{Z}}\int \d H\,\exp\bigg(-L\Tr\Big(V(H)\Big)\bigg)\,\exp\bigg(\sum_{n=1}^\infty \frac{G^n}{n}\Tr\Big(g^n\Big) \Tr(\Gamma\left(\mu-1/2\pm \i H^{1/2}\right)^n)\bigg)\nn\\&\qquad\qquad\qquad\qquad\qquad\qquad\qquad\qquad\qquad\qquad \delta_{i j}\,G\,\Tr\Big(\Gamma\left(\mu-1/2\pm \i H^{1/2}\right)e^{-\beta H}\Big)\,,\label{cont1bis}
\end{align}
with $\mathcal{Z}$ representing \eqref{iden1}. The second contribution \eqref{cont2} requires the large $L$ Gaussian approximation for the integral
\begin{equation}
    \int d U\, \Big(\C^\dagger\,U\,F(\Lambda)\,e^{-\beta \Lambda}\,F(\Lambda)\,U^\dagger\,\C\Big)_{j i}\, \exp\bigg(G\,\g^2\Tr\Big(\C\,\C^\dagger\,U\,F(\L)\,U^\dagger\Big)\bigg)\,.\label{HC}
\end{equation}
We need to account for Wick-contractions between the operator insertion and the exponential, in other words this calculation can be rewritten as
\begin{align}
    &\sum_{m=0}^\infty \frac{G^m}{m!}\,\g^{2 m}\,\bigg\langle\Big(\C^\dagger\,U\,F(\Lambda)\,e^{-\beta \Lambda}\,F(\Lambda)\,U^\dagger\,\C\Big)_{j i}\Tr\Big(\C\,\C^\dagger\,U\,F(\L)\,U^\dagger\Big)^m\bigg\rangle_\text{conn}\nn\\&\qquad\qquad\qquad\qquad\qquad\qquad\qquad\qquad\qquad\qquad\bigg\langle\exp\bigg(G\,\g^2\Tr\Big(\C\,\C^\dagger\,U\,F(\L)\,U^\dagger\Big)\bigg) \bigg\rangle\,,\label{aprox2}
\end{align}
the first line includes contractions between the insertion and the exponential, while the second line are the vacuum loops computed above. Summing over all large $L$ Gaussian Wick contractions one obtains
\begin{align}
    &\bigg\langle\Big(\C^\dagger\,U\,F(\Lambda)\,e^{-\beta \Lambda}\,F(\Lambda)\,U^\dagger\,\C\Big)_{j i}\Tr\Big(\C\,\C^\dagger\,U\,F(\L)\,U^\dagger\Big)^m\bigg\rangle_\text{conn}\nn\\&\qquad\qquad\qquad\qquad\qquad=m!\,\frac{1}{L^{m+1}}\Big(\Big(\C\,\C^\dagger\Big)^{m+1}\Big)_{j i} \Tr\Big(e^{-\beta H} \Gamma\left(\mu-1/2\pm \i H^{1/2}\right)^{m+2}\Big)\,.
\end{align}
Combining everything one finds that the second contribution \eqref{cont2} becomes
\begin{align}
    \bra{\psi_j}\ket{\psi_i}\supset\,&\frac{1}{\mathcal{Z}}\int \d H\,\exp\bigg(-L\Tr\Big(V(H)\Big)\bigg)\,\exp\bigg(\sum_{n=1}^\infty \frac{G^n}{n}\Tr\Big(g^n\Big) \Tr(\Gamma\left(\mu-1/2\pm \i H^{1/2}\right)^n)\bigg)\nn\\&\qquad\qquad\qquad\qquad\qquad \sum_{m=0}^\infty G^{m+2}\,\Big(g^{m+1}\Big)_{j i} \Tr\Big(e^{-\beta H}\Gamma\left(\mu-1/2\pm \i H^{1/2}\right)^{m+2}\Big)\,.\label{cont2bis}
\end{align}
Combining with \eqref{cont1bis}, we see that the matrix elements are indeed computed via the gravitational rules discussed in section \ref{sect:grav}; since we precisely recover \eqref{phiphiH}. The exponential within this $H$ integral, and the corresponding normalization with $1/\mathcal{Z}$ reflects the fact that we also include geometries where EOW particle loops are connected with the probe boundaries; as discussed below \eqref{phiphiH}. These calculations generalize in an obvious way to products of matrix elements like $\braket{\psi_j}{\psi_i}\braket{\psi_l}{\psi_k}$.

In summary, we have shown that all observables in the deformed matrix integral \eqref{model} are equivalent for weak coupling $\g$, with those of JT gravity with interacting EOW branes; establishing their duality.

\subsection{Strong coupling and non-random states}\label{sect:collapse}
Next we consider particular versions of our model, with propagator $1/G=\g+1$. For these, the partition function becomes\footnote{In this section we are only interested in the $C$ matrix integral and suppress the $H$ ensemble for presentation purposes.}
\begin{align}
    \mathcal{Z}=\int \d C\, \d C^\dagger \exp\bigg(-\Tr\Big(C^\dagger C\Big)-\g\Tr\Big(\Big(C-F(H)^{1/2}\C\Big)\Big(C^\dagger-\C^\dagger F(H)^{1/2}\Big)\Big)\bigg)\,.\label{modelbis}
\end{align}
This matrix model is very similar to the matrix model recently considered in \cite{drejorrit}; it interpolates between JT gravity with non-dynamical EOW branes \eqref{undef} at weak coupling $\g=0$
\begin{align}
    \mathcal{Z}=\int \d C\, \d C^\dagger \exp\bigg(-\Tr\Big(C^\dagger C\Big)\bigg)\,,\label{modelbis0}
\end{align}
and a gravity model with the matrix $C$ fixed to one member of the ensemble at strong coupling $\g=\infty$
\begin{align}
    \mathcal{Z}=\int \d C\, \d C^\dagger \exp\bigg(-\Tr\Big(C^\dagger C\Big)\bigg)\,\delta\Big(C-F(H)^{1/2}\C \Big)\delta\Big(C^\dagger-\C^\dagger F(H)^{1/2} \Big)\,.\label{modelbisinfty}
\end{align}
The stronger the interactions, the less random the matrix $C$, and the more realistic the quantum gravity model under consideration. This is one key lesson of this work, in these two dimensional models, realistic gravity systems appear to feature strong interactions/have strongly interacting interiors $g_{i j}\gg 1$. 

Equally important, the microscopic details of the theory, here represented by the non-random matrix $\C$, are encoded in the specific coupling constants $g_{i j}$ for the interior mode interactions, via \eqref{couplcoupl}. These coupling constants take typical values respecting the fact that $\C$ is a typical draw of the ensemble.

Concordantly, one can also interpret the completely random theory \eqref{modelbis0} as an ensemble of gravity theories, with the ensemble average over the specific coupling constants of the theory. The interpretation of \eqref{modelbis0} as bulk models with random couplings constants agrees with ideas of Coleman and company \cite{coleman1988black,giddings1988loss,giddings1989baby,Giddings:2020yes,Marolf:2020xie}. Here too, the theory with random couplings is ``simpler'' than that with fixed couplings.

In the most realistic models, where $\g=\infty$, the matrix elements of Hawking radiation \eqref{nondyn2} become
\begin{equation}
    \bra{\psi_j}\ket{\psi_i}=\Big(\C^\dagger\,F(H)\,e^{-\beta H}\,F(H)\,\C\Big)_{j i}\,.\label{nonrandom}
\end{equation}
This is obvious from \eqref{modelbisinfty}, but can also be seen explicitly in \eqref{cont1} and \eqref{cont2}. The contribution of the first Wick contraction \eqref{cont1} vanishes, because in this double scaling regime the propagator vanishes $G=0$. The contribution of the second Wick contraction survives, because in the same regime $G^2 \g^2 =1$.

This extends to products of matrix elements, EOW branes without any interaction vertices on them are suppressed because $G=0$; for the product of two matrix elements one obtains for example
\begin{equation}
    \bra{\psi_j}\ket{\psi_i}\bra{\psi_l}\ket{\psi_k}=\Big(\C^\dagger\,F(H)\,e^{-\beta H}\,F(H)\,\C\Big)_{j i}\Big(\C^\dagger\,F(H)\,e^{-\beta H}\,F(H)\,\C\Big)_{l k}\,,\label{fixedtwo}
\end{equation}
where $H$ remains random but $\C$ is non-random. The non-randomness of $\C$ does not imply that \eqref{fixedtwo} is numerically close to factorizing, or to the matrix elements in a completely non-random gravity theory. Since $H$ remains random there remains large correlation between two density matrix elements. This is largely because of the random unitaries $U$; not so much the more common eigenvalue correlation.\footnote{These random unitaries did not play any role for the non-dynamical theory \eqref{model}, because its action is $U(L)$ invariant. However this $U(L)$ invariance is broken by the deformation \eqref{modelbis} because $\C$ is some fixed matrix not proportional to the identity.}

The bulk gravity description of the matrix elements of Hawking radiation in realistic incarnations of these two dimensional quantum gravity models, described by non-random matrices $\H$ and $\C$, involves at minimum these interior mode interactions.

On top of that, these involve extra ingredients that capture the non-random matrix $\H$. What these ingredients are is an orthogonal question that goes beyond our current scope; progress has been made in this regard in \cite{Blommaert:2019wfy,Blommaert:2020seb,drejorrit,Saad:2021rcu}, see section \ref{sect:disc}. Regardless of the specific details of those extra ingredients, the result is some gravity theory whose matrix elements are good-old-fashioned non-random numbers
\begin{equation}
    \bra{\psi_j}\ket{\psi_i}=\Big(\C^\dagger\,F(\H)\,e^{-\beta \H}\,F(\H)\,\C\Big)_{j i}\,.\label{nonrandombis}
\end{equation}
The structure in these numbers for a typical draw of the ensemble deserves some attention.

\section{Matrix elements without ensemble average}\label{sect:discrete} 
In this section we examine the consequences of collapsing the theory to a single member of the ensemble, by studying numerical features of the matrix elements \eqref{nonrandombis}, and how they reproduce the Page curve.

Throughout this section, we work in a microcanonical ensemble centered around $\E$, with $e^{\Ss}$ states. Since $F(\H)$ is approximately constant for eigenvalues of $\H$ within the microcanonical window we have
\begin{equation}
     \bra{\psi_j}\ket{\psi_i}=F(\E)^2\sum_{\alpha=1}^{e^{\Ss}} {\textsf{C}_{0}}^*_{\a j}\,{\textsf{C}_{0}}^{\,}_{\a i}\,.
\end{equation}
The kernel $F(\E)$ drops out of the normalized density matrix and all derivative quantities like the Renyi entropies and the von Neumann entropy, this means we can immediately drop it and continue with
\begin{equation}
     \bra{\psi_j}\ket{\psi_i}=\sum_{\alpha=1}^{e^{\Ss}} {\textsf{C}_{0}}^*_{\a j}\,{\textsf{C}_{0}}^{\,}_{\a i}\,.\label{modmod}
\end{equation}

Crucially, the matrix $\C$ must be interpreted as typical representative of the undeformed $C$ ensemble \eqref{modelbis0}. This means the entries of $\C$ are complex numbers with typical norm-squared one.\footnote{Fixing to non-typical members is physically non-sensible, since those systems are not even remotely accurately described by the ensemble to begin with.}

Formula \eqref{modmod} is a toy model for the matrix elements of evaporating black holes in quantum gravity. The ensemble is an incredibly powerful tool that simplifies calculations and presents a simpler effective picture, but is has also caused confusion. Mistaking the ensemble for proper quantum mechanics, one apparently finds that the density matrix is maximally mixed, as in Hawking's calculation \cite{Hawking:1975vcx}.

In realistic theories, represented by the toy model \eqref{modmod}, the density matrix is \emph{not} maximally mixed. The complex matrix $\C$ has dimensions $e^{\Ss}\times k$, and concordantly the rank of the matrix is upper bound by both $k$ and $e^{\Ss}$ \cite{Marolf:2020xie}; this suffices to understand the Page curve. The point is that when we are ignorant about the microstructure of our system, that we believe the states $\ket{\psi_i}$ are linearly independent; where in reality there are equivalence relations between them, there are null states. The remaining question is then how microstructure gets realized in the bulk, this work has been a step in that direction.\footnote{Factorization is another interesting question, this remains geometrically nontrivial even in microscopic realizations; we have not investigated that here \cite{Maldacena:2004rf,Saad:2019lba,Blommaert:2019wfy,Penington:2019kki,Blommaert:2020seb,Stanford:2020wkf,Saad:2021rcu,Iliesiu:2021are,Saad:2021uzi}.}

We plotted the density matrix in Fig. \ref{fig:rho}, the purity in Fig. \ref{fig:purity}, and the von Neumann entropy in Fig. \ref{fig:entropy} for one single matrix $\C$. Comparison with the averaged answers confirms the latter are self-averaging. In the remainder of this section we explain analytically why \eqref{modmod} produces these plots.

\subsection{Density matrix}\label{sect:31}
\begin{figure}
    \centering
    \begin{equation}
        \text{Re}(\rho_{i j})\,\, \raisebox{-22mm}{\includegraphics[scale=.4]{Figures/rho_diag1.png}}\quad\quad\raisebox{-22mm}{ \includegraphics[scale=.4]{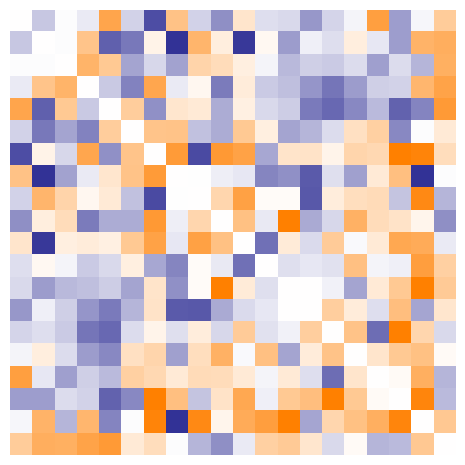}} \nn
    \end{equation}
    \caption{Normalized density matrix with $k=20$ and $e^{\Ss}=50$ (left). Orange is positive and blue negative. The diagonal is removed (right) and the intensity is rescaled to probe typical sizes $1/k\, e^{\Ss/2}$. This reproduces similar plots obtained within SYK, with notably a different realization for the density matrix than \eqref{eqn:rho_singlemember} \cite{Stanford:2020wkf}.}
    \label{fig:rho}
\end{figure}

The unnormalized density matrix of radiation is plotted in  Fig. \ref{fig:rhobis} and Fig. \ref{fig:rho}, and reads explicitly
\be \label{eqn:rho_singlemember}
\rho=\sum_{i,j=1}^{k}\bra{\psi_j}\ket{\psi_i} \ket{i}\bra{j} = \sum_{i,j=1}^{k} \sum_{\alpha=1}^{e^{\Ss}} {\textsf{C}_{0}}^*_{\a j}\,{\textsf{C}_{0}}^{\,}_{\a i} \ket{i}\bra{j}\,.
\ee

To understand this figure, let us estimate the typical size of the matrix elements. The entries of $\C$ are complex numbers with typical norm squared one, which means that the diagonal matrix elements
\begin{equation}
    \bra{\psi_i}\ket{\psi_i}=\sum_{\alpha=1}^{e^{\Ss}} {\textsf{C}_{0}}^*_{\a i}\,{\textsf{C}_{0}}^{\,}_{\a i}\,\label{34}
\end{equation}
are real numbers with typical size $e^{\Ss}$, because they are the sum of $e^{\Ss}$ real numbers of typical size one. 

The off-diagonal matrix elements 
\begin{equation}
    \bra{\psi_j}\ket{\psi_i}=\sum_{\alpha=1}^{e^{\Ss}} {\textsf{C}_{0}}^*_{\a j}\,{\textsf{C}_{0}}^{\,}_{\a i}\,\label{36}
\end{equation}
are complex numbers with a ``random'' phase, and typical amplitude of size $e^{\Ss/2}$. This is because they are the sum of $e^{\Ss}$ complex numbers with random phases and typical amplitudes of size $1$, combined they are to be interpreted as a ``random'' walk in the complex plane with unit step length and $e^{\Ss}$ steps\footnote{Random is quoted because there is nothing uncertain about the number in \eqref{36}, ``typical'' might be more appropriate.}. The typical radial distance traveled by this random walk is the square root of the number of steps $e^{\Ss/2}$.

The off-diagonal terms correspond to sums over random complex numbers that do not constructively interfere, the phases in \eqref{34} align but those in \eqref{36} do not, hence they give small contributions relative to the diagonal terms as long as $e^{\Ss}\gg 1$.

Alternatively one could directly estimate the norm squared for the off-diagonal matrix elements
\begin{equation}
    \abs{\bra{\psi_j}\ket{\psi_i}}^2=\sum_{\alpha_1,\alpha_2=1}^{e^{\Ss}} {\textsf{C}_{0}}^*_{\a_1 j}\,{\textsf{C}_{0}}^{\,}_{\a_1 i}\,{\textsf{C}_{0}}^*_{\a_2 i}\,{\textsf{C}_{0}}^{\,}_{\a_2 j}\,\,.\label{37}
\end{equation}
The most sizable contribution comes from the terms with $\a_1=\a_2$. These are $e^{\Ss}$ real numbers of typical size one, adding up to give some real contribution of typical size $e^{\Ss}$. There are other terms that pair up to form real numbers, combining terms where we exchange $\a_1$ and $\a_2$. The typical size of the sum of such two terms is $\sqrt{2}$; but there is typically the same number of terms with overall positive sign, and overall negative sign. Therefore these are subleading. Crucially, the leading contributions always come from terms where the phases precisely cancel.

Typical values of matrix elements are most efficiently estimated by using the ensemble average, this is what typicality means. In the ensemble averaged description, one indeed computes
\begin{equation}
    \bra{\psi_i}\ket{\psi_j}\overset{\text{aver}}{=}\delta_{i j}\,e^{\Ss}\,,\quad \abs{\bra{\psi_j}\ket{\psi_i}}^2\overset{\text{aver}}{=}\delta_{i j}\,e^{2\Ss}+e^{\Ss}\,,\label{avav}
\end{equation}
reproducing the above typical estimates. These formulas also highlight the discussion below \eqref{modmod}. The leading order approximation for the density matrix is the maximally mixed diagonal Hawking state; but there are subleading non-self-averaging contributions of order $e^{\Ss/2}$ in the unnormalized density matrix
\be
\rho=e^{\Ss}\sum_{i=1}^k\ket{i}\bra{i} +\mathcal{O}(e^{\Ss/2})\,.
\ee
We cannot emphasize enough that these subleading corrections can, and do save unitary \cite{Papadodimas:2012aq,Papadodimas:2013kwa,Stanford:2020wkf,Penington:2019kki}. We will show in the following section that these subleading corrections are important for observables. Their effect is to cause linear relations between the different vectors $\ket{\psi_i}$
\begin{equation}
    \sum_{i=1}^k a_i \ket{\psi_i}=0\,.
\end{equation}
The off-diagonal matrix elements do not need to be leading order for such linear relations to exist. The density matrix \eqref{eqn:rho_singlemember} shown in Fig. \ref{fig:rho} and Fig. \ref{fig:rhobis}, is an extremely concrete example of how this happens. The leading order approximation is diagonal, but by construction the dimension of the span of the $\ket{\psi_i}$, which equals the rank of $\C$; is upper bound by both $k$ and $e^{\Ss}$. The off-diagonal matrix elements are therefore responsible for the Page transition at late times.

Circling back to the gravitational picture of section \ref{sect:grav}, notice in \eqref{couplcoupl} that the typical off-diagonal couplings $g_{i j}$ are subleading compared to the diagonal couplings $g_{i i}$. This intuitively explains why the diagonal matrix elements remain bigger than the off-diagonal ones, even in the interacting theory.

\subsection{Higher moments}\label{sect:higher}
Next we examine how calculations of various entropies in a typical member of the ensemble are consistent with the answers given by the replica wormhole calculations of \cite{Penington:2019kki}. We will find that off-diagonal matrix elements constructively add up and give large contributions to observables, corresponding to the effects of replica wormhole geometries. 

\begin{figure}
    \centering
    \includegraphics[scale=0.5]{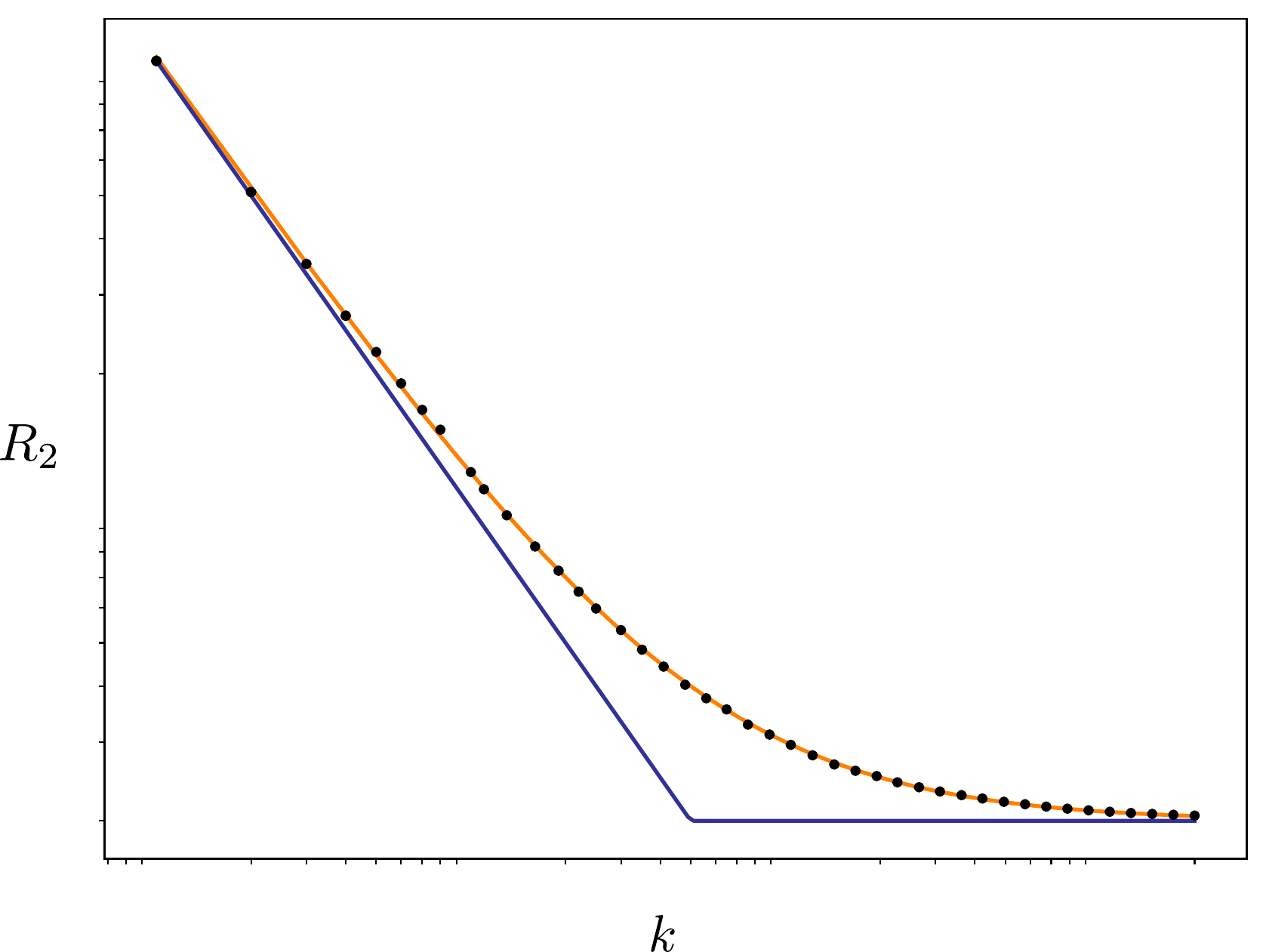}
    \caption{Purity $R_2$ as function of $k$ with $e^{\Ss}=50$, showing \eqref{purpu} (black dots) and the planar approximation $1/k+1/L$ (orange), log-log representation.}\label{fig:purity}
\end{figure}

The simplest entropy observable is the purity, which is plotted in  Fig. \ref{fig:purity} and reads explicitly
\be
R_2=\frac{1}{\Tr(\rho)^2}\sum_{i,j=1}^k\bra{\psi_j}\ket{\psi_i}\bra{\psi_i}\ket{\psi_j}=\frac{1}{\Tr(\rho)^2}\sum_{i,j=1}^k \sum_{\alpha_1,\alpha_2=1}^{e^{\Ss}} {\textsf{C}_{0}}^*_{\a_1 j}\,{\textsf{C}_{0}}^{\,}_{\a_1 i}\,{\textsf{C}_{0}}^*_{\a_2 i}\,{\textsf{C}_{0}}^{\,}_{\a_2 j}\,,\label{purpu}
\ee
here $\Tr(\rho)$ is computed using \eqref{eqn:rho_singlemember}. As explained below \eqref{37} the leading contributions come from terms in the sums where two phases align, meaning that complex matrix elements of $\C$ pair up as
\begin{equation}
    \wick[wickcolor=ourorange]{\c1 {\textsf{C}_{0}}^*_{\a i}\,\c1{\textsf{C}_{0}}^{\,}_{\a i}}\overset{\text{est}}{=}1\,.\label{pair}
\end{equation}
In the purity summation \eqref{purpu}, this happens if $i=j$ and or $\a_1=\a_2$. Using the leading order estimate $\Tr(\rho)=k\,e^{\Ss}$ one then estimates the typical purity\footnote{These orange ``Wick contractions'' represent the ways that complex matrix elements can pair up in real numbers \eqref{pair}.}
\begin{align} 
    R_2&\overset{\text{est}}{=}\frac{1}{k} + \frac{1}{e^{\Ss}} + \frac{1}{ k\, e^{\Ss}}\,\text{Re}( e^{i\phi})\label{r222}\\&=\frac{1}{k^2}\sum_{i,j=1}^k \sum_{\alpha_1,\alpha_2=1}^{e^{\Ss}} \wick[wickcolor=ourorange]{\c1{\textsf{C}_{0}}^*_{\a_1 j}\,\c1{\textsf{C}_{0}}^{\,}_{\a_1 i}}\,\wick[wickcolor=ourorange]{\c1{\textsf{C}_{0}}^*_{\a_2 i}\,\c1{\textsf{C}_{0}}^{\,}_{\a_2 j}}+\frac{1}{k^2}\sum_{i,j=1}^k \sum_{\alpha_1,\alpha_2=1}^{e^{\Ss}} \wick[wickcolor=ourorange]{\c2{\textsf{C}_{0}}^*_{\a_1 j}\,\c1{\textsf{C}_{0}}^{\,}_{\a_1 i}\,\c1{\textsf{C}_{0}}^*_{\a_2 i}\,\c2{\textsf{C}_{0}}^{\,}_{\a_2 j}}+\dots\nn
\end{align}

The first contribution in \eqref{r222} comes from terms with $i=j$, corresponding with diagonal matrix elements or the disk geometries in \eqref{pur}. By itself these diagonal terms claim the purity of the radiation decreases forever as $1/k$. However, there is another important contribution from the terms with $\alpha_1=\a_2$. This term counts the norm squared of the non-self-averaging fluctuations in \emph{all} of the matrix elements $\rho_{i j}$, represented by the final contribution in \eqref{avav}. The noise in the off-diagonal elements constructively interferes when computing the purity, this corresponds with the replica wormhole in \eqref{pur} \cite{Penington:2019kki,Stanford:2020wkf}.

There are $k^2$ matrix elements with fluctuations, whereas there are only $k$ diagonal matrix elements whose self-averaging behavior gives the Hawking answer. For late enough times $k$ the non-self-averaging fluctuations win over the self-averaging terms and cause the Page transition \cite{Papadodimas:2012aq,Papadodimas:2013kwa,Penington:2019kki,Stanford:2020wkf}. Here this is represented by the second contribution in \eqref{r222} winning over the first one \cite{Penington:2019kki}, see also Fig. \ref{fig:purity}.

These corrections actually become much more important when constructing a more standard Page curve, keeping the dimension of the Hilbert space of the total system fixed $k\, e^{\Ss}=d$, whilst increasing $k$. For extremely old black holes $e^{\Ss}\,\,\propto\,\, 1$, the off-diagonal terms are not small anymore; the size of the non-self-averaging fluctuations in \eqref{avav} becomes comparable to the size of the signal itself, as in Fig. \ref{fig:rhobis}. It is hence not true that \emph{small} corrections save unitarity \cite{Papadodimas:2012aq,Papadodimas:2013kwa}, \emph{large} off-diagonal matrix elements do.

The last term in \eqref{r222} estimates the size of non-self-averaging fluctuations, it comes from terms in the sum where $a_1\neq \a_2$ and $i\neq j$. Each term in this sum is a ``random'' complex number with typical norm squared one. There are roughly $k^2\, e^{2\Ss}$ such terms, therefore the sum represents a ``random'' walk that travels a typical radial distance $k\,e^{\Ss}$. For large $k$ and $e^{\Ss}$ these fluctuations are small, meaning that the purity is self-averaging. Ensemble averages accurately compute entropies, but not matrix elements.

This generalizes to the other moments $R_n=\Tr(\rho^n)/\Tr(\rho)^n$
\begin{equation}
    R_n=\frac{1}{\Tr(\rho)^n}\sum_{i_1\dots i_n=1}^k\bra{\psi_{i_1}}\ket{\psi_{i_2}}\dots \bra{\psi_{i_n}}\ket{\psi_{i_1}}=\frac{1}{\Tr(\rho)^n}\sum_{i_1\dots i_n=1}^k \sum_{\alpha_1\dots \alpha_n=1}^{e^{\Ss}} {\textsf{C}_{0}}^*_{\a_1 i_1}\,{\textsf{C}_{0}}^{\,}_{\a_1 i_2}\dots{\textsf{C}_{0}}^*_{\a_n i_n}\,{\textsf{C}_{0}}^{\,}_{\a_n i_1}\,\,.
\end{equation}
For $n>2$ there are more ways the matrix elements $\C$ can pair with their complex conjugates \eqref{pair}, corresponding with the sum over Wick contractions in the ensemble averaged calculation, and with the different replica wormhole geometries in the model of \cite{Penington:2019kki}. For example for $n=3$, $4$, one estimates
\begin{equation}
    R_3\overset{\text{est}}{=}\frac{1}{k^2}+\frac{3}{k\,e^{\Ss}}+\frac{1}{e^{2\Ss}}\,,\quad R_4\overset{\text{est}}{=}\frac{1}{k^3}+\frac{6}{k^2\,e^{\Ss}}+\frac{6}{k\,e^{2\Ss}}+\frac{1}{e^{3\Ss}}\,.\label{r3r4}
\end{equation}
Here we are no longer tracking the non-self-averaging fluctuations and we dropped terms of subleading order in either $k$ or $e^{\Ss}$. This corresponds with using the planar approximation in replica wormholes \cite{Penington:2019kki}.

\subsection{Entropy and planar resummation}\label{sect:entropy}
\begin{figure}
    \centering
    \includegraphics[scale=0.5]{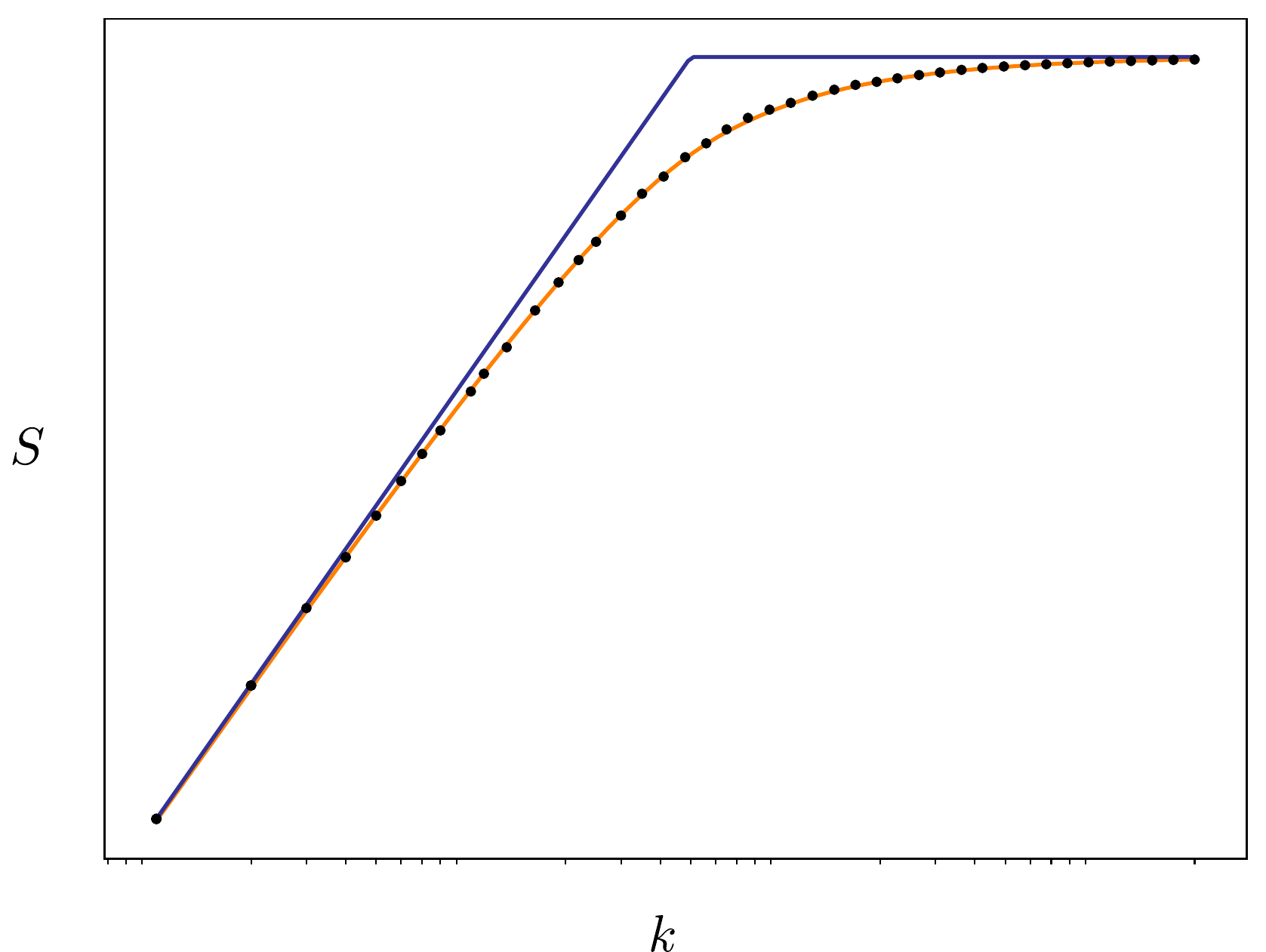}
    \caption{Entropy $S$ as function of $k$ with $e^{\Ss}=50$, showing \eqref{vne} (black dots) and the planar approximation \eqref{eqn:planar_entropy} (orange).}
    \label{fig:entropy}
\end{figure}

Using \eqref{eqn:rho_singlemember} we can directly compute the von Neumann entropy, by literally computing the log of matrix, using henceforth the normalized version of the density matrix \eqref{eqn:rho_singlemember}
\begin{equation}
    S=-\Tr(\rho \log \rho)\,.\label{vne}
\end{equation}
This entropy is plotted in Fig. \ref{fig:entropy}. We now reproduce this figure using the planar approximation for the moments discussed above, by explicitly applying the replica trick
\begin{equation}
    S=-\partial_n R_n\big\rvert_{n=1}\,.\label{reptrick}
\end{equation}
This is nontrivial as it requires finding the unique analytic expression for the Renyis as function of $n$.

The first difficulty is computing $R_n$ for arbitrary positive integer $n$ by summing all planar diagrams. This is in principle a difficult counting problem, the key to solving this implicitly was discussed in \cite{Penington:2019kki}, following \cite{speicher2009free,cvitanovic1981planar}. To obtain the Renyi entropies as function of $n$ one should instead compute a generating function, for example the resolvent of the density matrix
\begin{equation}
R(\lambda)=\Tr(\frac{1}{\lambda -\rho})=\frac{k}{\lambda}+\sum_{n=1}^{\infty} \frac{1}{\lambda^{n+1}}\Tr(\rho^n)\,.\label{resresres}
\end{equation}
Its Taylor expansion around infinite $\lambda$ encodes the moments $R_n$, and hence also the Renyi entropies.
This expansion, along with the structure of the planar geometries that contribute, makes it possible to write down a Schwinger-Dyson equation for $R(\lambda)$; in our microcanonical setup this becomes \cite{Penington:2019kki}
\begin{equation}
R(\lambda)^{2}+\left(\frac{e^{\Ss}-k}{\lambda}-k\,e^{\Ss}\right) R(\lambda)+\frac{k^{2}\,e^{\Ss}}{\lambda}=0.
\end{equation}
According to \eqref{resresres} the solution must behave as $k/\lambda$ near $\lambda=\infty$; furthermore recognizing the generating functional of Gegenbauer polynomials one then obtains the unique solution
\begin{align}
R(\lambda) &= \frac{1}{2} k\,e^{\Ss}-\frac{1}{2}\frac{e^{\Ss}-k}{\lambda}-\frac{1}{2}\left(-\frac{4 k^2 \,e^{\Ss}}{\l}+\left(\frac{e^{\Ss}-k}{\lambda}-k\,e^{\Ss}\right)^2\right)^{1/2}\\&=\frac{k}{\lambda} -\frac{k L}{2} \sum_{n=2}^{\infty}\frac{1}{\lambda^n} \left(\frac{L-k}{k L}\right)^{n} C_{n}^{(-1 / 2)}\left(\frac{L+k}{L-k}\right)\,.
\end{align}
Using the relation between Gegenbauer polynomials and Jacobi polynomials,\footnote{\begin{equation}
C_{n}^{(-1/2)}(x) = \frac{-2}{n-1} P_{n}^{(-1, -1)}(x)= -\frac{2}{n-1} \sum_{s=0}^{n} \binom{n-1}{n-s} \binom{n-1}{s} \left(\frac{x-1}{2}\right)^{s}\left(\frac{x+1}{2}\right)^{n-s}
\end{equation}} we can rewrite this into
\begin{equation}
    R(\lambda)=\frac{k}{\l}+\sum_{n=2}^\infty \frac{1}{\l^n}\sum_{s=0}^n\frac{\Gamma(n-1)\Gamma(n)}{\Gamma(n-s+1)\Gamma(n-s)\Gamma(s+1)\Gamma(s)}\,k^{(s+1-n)}\,e^{(1-s)\Ss}\,.
\end{equation}
From the expansion coefficients one therefore finds the moments\footnote{This formula was also recently derived in \cite{Kudler-Flam:2021rpr,Kudler-Flam:2021alo}.} 
\begin{align}
    R_n&=\sum_{s=0}^{n+1}\frac{\Gamma(n)\Gamma(n+1)}{\Gamma(n-s+2)\Gamma(n-s+1)\Gamma(s+1)\Gamma(s)}\,k^{(s-n)}\,e^{(1-s)\Ss}\nn\\&=\sum_{p=0}^{n+1}\frac{\Gamma(n)\Gamma(n+1)}{\Gamma(n-p+2)\Gamma(n-p+1)\Gamma(p+1)\Gamma(p)}\,k^{(1-p)}\,e^{(p-n)\Ss}\,,\label{sumsumss}
\end{align}
where changing coordinates as $p=n+1-s$ highlights the symmetry under exchange of $k$ and $e^{\Ss}$. This reproduces the correct answer for any integer $n$. One easily checks the simplest cases \eqref{r222} and \eqref{r3r4}.

The second, we believe generally less appreciated difficulty, is finding a unique analytic continuation of this formula away from integer $n$. Usually, one hopes that one obvious analytic continuation presents itself; however here there are two obvious and inequivalent options. As $1/\Gamma(n+1-s)=0$ for $s>n+1$ one can extend the range of the first sum from $s=0$ to $s=\infty$. Via the same argument one may extend the second sum from $p=0$ to $p=\infty$, for positive integers $n$.\footnote{This corresponds with extending the first sum from $s=-\infty$ to $s=n+1$, doing both does not yield a convergent sum.} Both procedures give a hypergeometric function,\footnote{Hypergeometric functions are defined as semi-infinite sums, tread carefully with Mathematica here.} these agree on the positive integers but \emph{disagree} elsewhere; resulting in \emph{two} possible analytic continuations (as function of $n$)
\begin{equation}
    R_n=\frac{1}{k^{(n-1)}} \,_2F_1(-n,1-n,2,k/e^{\Ss})\quad \text{or}\quad R_n=\frac{1}{e^{(n-1)\Ss}} \,_2F_1(-n,1-n;2;e^{\Ss}/k)\,,\label{choice}
\end{equation}
which are swapped when exchanging $k$ and $e^{\Ss}$. Then which of these is the correct analytic continuation? We need a theorem that specifies uniqueness of analytic continuation, given data at the positive integers.

The only such theorem that we know of is due to Carlson, see \cite{Engelhardt:2020qpv}. If there is a function $f(z)$ that is analytic for $\text{Re}(z)\geq 0$, that takes assigned values $f_n$ on the positive integers, that grows exponentially slower than $\sin(\pi z)$ for imaginary $z$, and no faster than exponential elsewhere; then this function $f(z)$ is the unique one with these properties. 

Conversely, the data $f_n$ is insufficient to uniquely specify an analytic function $f(z)$ without further constraints, and Carlson's theorem gives the necessary constraints that uniquely select one function. It is not a priori obvious why the moments $R(z)$ should satisfy these constraints, but Carlson's theorem proves that if they do not, then there is no unique function $R(z)$; and therefore no unique von-Neumann entropy. That is clearly nonphysical, therefore we conclude that $R(z)$ \emph{must} satisfy Carlson's theorem.\footnote{Otherwise we can just add the function $c \sin(\pi z)$ to the moments with arbitrary $c$, which contributes $\pi c$ to the entropy.}

The first function in \eqref{choice} satisfies this theorem as function of $n$ when $k<e^{\Ss}$, however it grows too quickly on the negative imaginary axis when $k>e^{\Ss}$; the second function in \eqref{choice} satisfies the theorem for the same reason only when $k>e^{\Ss}$. Therefore the unique analytic continuation in $n$ is\footnote{The Heavisides also conveniently save us from evaluating the hypergeometric on the branchcut it has in the last variable.}
\begin{equation}
R_n=\frac{1}{k^{n-1}} \,_2F_1(-n,1-n,2,k/e^{\Ss})\,\theta(k<e^{\Ss})+\frac{1}{e^{(n-1)\Ss}} \,_2F_1(-n,1-n;2;e^{\Ss}/k)\theta(k>e^{\Ss})\,.
\end{equation}
With this, one computes the entropy using the replica trick \eqref{reptrick}. This reproduces the Page curve \cite{Page:1993df}\footnote{Analytically, using the representation of the hypergeometric as infinite sums \eqref{sumsumss} from $s=0$ to $s=\infty$, one takes the derivatives of the Gamma functions, then uses the poles and residues of the Gamma and Digamma functions to prove that only one term in the sum contributes. It is crucial that in the analytic continuation the sums are semi-infinite, otherwise the derivative is not well defined.}
\begin{equation}
    S=(\log(k)-k/2e^{\Ss})\,\theta(k<e^{\Ss})\,+\,(\Ss-e^{\Ss}/2k)\,\theta(k>e^{\Ss})\,.\label{eqn:planar_entropy}
\end{equation}
This agrees excellently with a direct calculation of the entropy \eqref{vne} using our density matrix \eqref{eqn:rho_singlemember}, see Fig. \ref{fig:entropy}. It is not surprising that the Page curve is self-averaging, the point is that we have produced it using the density matrices in Fig. \ref{fig:rhobis}, confirming the claims made about off-diagonal matrix elements in section \ref{sect:higher}. More importantly, we have given a gravitational interpretation for these matrix elements in section \ref{sect:grav}.

\section{Concluding remarks}\label{sect:disc}
In this work we made progress towards understanding the bulk gravity dual to one quantum system. We investigated how the density matrix elements of evaporating black holes are computed in non-random gravity theories, and in particular what explain small off-diagonal density matrix components.

For this we investigated an enrichment of the model of Pennington, Shenker, Stanford and Yang, by including dynamics for EOW branes; namely brane flavor changing interaction vertices and loops of EOW branes; and discovered a dual description as a deformed matrix integral
\begin{align}
    \mathcal{Z}=\int \d C\, \d C^\dagger \exp\bigg(-\Tr\Big(C^\dagger C\Big)-\g\Tr\Big(\Big(C-F(H)^{1/2}\C\Big)\Big(C^\dagger-\C^\dagger F(H)^{1/2}\Big)\Big)\bigg)\,,
\end{align}
where the coupling constants for the interaction vertices are related to the matrix deformation as
\begin{equation}
    g_{i j}=\g^2 \frac{1}{L}\Big(\C^\dagger\,\C\Big)_{j i}\,.\label{dictionarrry}
\end{equation}

For increasing values of the coupling constant $\g$, and hence $g_{i j}$, the random matrix $C$ gets gradually fixed to a non-random matrix $\C$. This means that the interior states $\ket{\psi_i}$ are becoming less random.

Our main conclusions are the following.
\begin{enumerate}
    \item Strong interactions behind the horizon are essential for understanding the microstructure of matrix elements of evaporating black holes, within this simple model.
    \item The microscopic details of the theory, here represented by the non-random matrix $\C$, are encoded in gravity as coupling constants for interior interactions.
    \item Large amounts of tiny off-diagonal matrix elements eventually overtake the bigger diagonal matrix elements, these correspond with replica wormholes and cause the Page curve transition.
    \item For nearly evaporated black holes off-diagonal matrix elements are large, and the state approaches some pure state as required for unitarity.
\end{enumerate}

In a gravity model where the density matrix of Hawking radiation is described by \eqref{nonrandom}
\begin{equation}
    \rho_{i j}=\Big(\C^\dagger\,F(H) e^{-\beta H} F(H)\, \C\Big)_{j i}\,,
\end{equation}
one immediately sees that there are nontrivial off-diagonal matrix elements, without having to compute their variance. The raison d'\^{e}tre for the simplified ensemble averaged gravity theories is they are simple to compute with, this is the whole philosophy behind random matrix theory \cite{mehta2004random,Haake:1315494}. They were never meant to describe the microstructure of individual systems, we should not forget this. Random matrices are sufficiently smart to understand \emph{that} the off-diagonal matrix elements are nonzero. But they are not a microscopic description of the theory, where we can actually understand \emph{why} they are nonzero. The real universe is clearly not an ensemble average; no one would claim Navier-Stokes is the fundamental description of fluids, neither are random matrices the fundamental description of the bulk.

Simple effective description like JT gravity with non-dynamical EOW branes, Brownian motion, and pure Einstein-Hilbert gravity in higher dimensions \cite{Maldacena:2004rf,Cotler:2020ugk} are ensembles. However, real fundamental description like deformed JT gravity \cite{drejorrit} with dynamical EOW particles, atoms, and full-fledged string theory \cite{Eberhardt:2020bgq} are factorizing and unitary quantum systems without ensembles.

That being said, ensemble averaged descriptions are clearly extremely useful, precisely because they corresponds with simple gravitational duals; those simple duals suffice for many calculations.

We end this work with several comments, first and foremost about higher dimensional implications.

\subsection*{General lessons}
We believe our findings are evidence that strong interactions in the interior will generically be important to capture the microstructure of higher dimensional black holes. These interactions factorize (replica) wormholes, because they collapse the ensemble, and because without the ensemble everything factorizes. Knowing how to calculate the microscopic out-state of the radiation is equivalent to understanding how to factorize (replica) wormholes.

In our model factorization is less geometrically obvious than is the case with eigenvalue correlation \cite{Blommaert:2019wfy,Blommaert:2020seb}, where there is some exclusion rule and concordantly a diagonal $=$ cylinder identity \cite{Saad:2021uzi}. In this setup, when we calculate in gravity the product of two matrix elements, there is the replica wormhole; but also other connected components, from both matrix elements connecting to the D-brane partition function (or EOW loops). They can both be connected to EOW loops via wormholes, or via the nonlocal interactions discussed below. Since the replica wormholes are not related to eigenvalue correlation, we believe that the nonlocal interactions might be the key. Somehow the replica wormholes should then be canceled by nonlocal interactions between different copies, restoring factorization. This must happen, because the ensemble is collapsed, nevertheless it would be interesting to make this more precise.

\begin{figure}
    \begin{equation}
        \bra{\psi_j}\ket{\psi_i}=\delta_{i j}\quad\raisebox{-13mm}{\includegraphics[width=22mm]{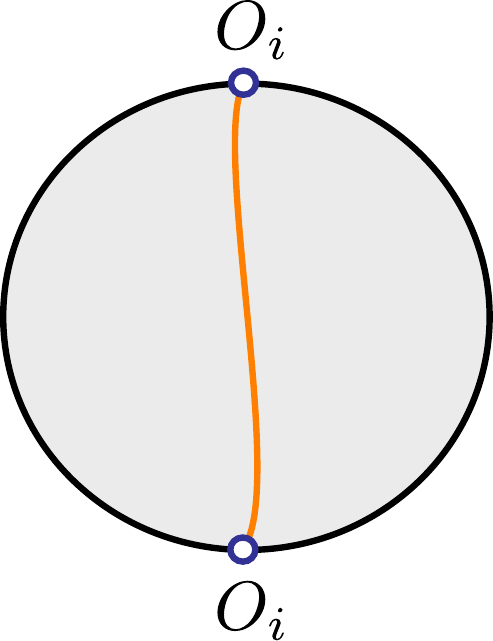}}\quad +\quad \raisebox{-13mm}{\includegraphics[width=22mm]{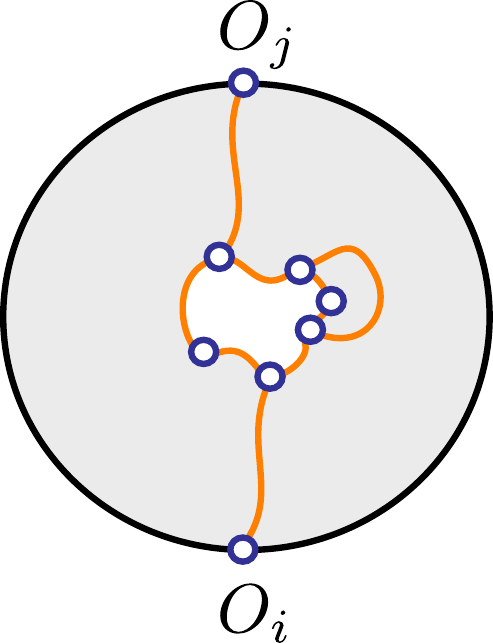}}\nn
    \end{equation}
    \caption{Picture for (two-sided) matrix elements that generalizes to higher dimensions. One could imagine creating orthogonal states by preparing states with particles of different flavors. In microscopic models the particles could interact heavily in the interior, perhaps close to the singularity, resulting in off-diagonal matrix elements. This conclusion is similar to the effective half-wormholes for eigenvalue correlation in \cite{Saad:2021rcu,Saad:2021uzi} and also to the picture of \cite{Blommaert:2019wfy,Blommaert:2020seb,drejorrit}. Strong interior interactions encode microstructure. This also applies in \cite{Almheiri:2019qdq}.}\label{fig:general}
\end{figure}

This picture we obtain here is, perhaps surprisingly, morally related to the one advocated in \cite{Saad:2021rcu,Saad:2021uzi}, where they discuss an effective description for (eigenvalue) microstructure. The information about that microstructure is located in the interior, perhaps even near the singularity. We have a different setup here, and are describing different aspects of black hole microstructure, namely the out-state of radiation. Nevertheless the overall lesson is similar: strong interior interactions encode microstructure. This is also one possible interpretation of \cite{Blommaert:2019wfy,Blommaert:2020seb,drejorrit}, which gives a precise description of eigenvalue microstructure. See Fig. \ref{fig:general}

The EOW branes studied in this paper are behind the horizon \cite{Kourkoulou:2017zaj,Gao:2021uro}. But one could consider an alternative version of this same model, with negative energy dynamical branes modeling random states, but these are outside of the horizon \cite{Gao:2021uro}. Their coupling constants would still encode microstructure, but that information would now be outside. Perhaps this is a sensible toy model for fuzzballs; it would be interesting to connect that literature better with the current developments concerning wormholes and ensembles \cite{Bena:2007kg,Mathur:2009hf,Mathur:2014nja}. 

Concerning the dependence of the couplings on $\C$ we believe the generalized picture is the following. Consider a UV complete theory of quantum gravity, like string field theory. These theories are probably rather unique and special, concordantly the couplings between the matter fields in the spectrum would take rather specific values. One could imagine integrating out most fields in the spectrum, leaving some ``simplified'' model with fewer fields; like dilaton gravity with EOW branes. 

However integrating out the matter fields would leave its imprint, it would not leave something nice and simple like ordinary JT gravity, with non-dynamical EOW branes. Rather, one would obtain some highly deformed JT gravity with complicated dilaton gravity interactions, and dynamical EOW branes. The details about the UV microstructure would get imprinted in all these interactions, see also \cite{drejorrit}.

Concerning the off-diagonal matrix elements there should obviously be a generalization to arbitrary black holes. States describing the interior of black holes should acquire nontrivial overlaps as the black holes grow old, to ultimately restore unitarity. 

The mechanism by which these states become equivalent is an important open question, our results suggest that strong interior interactions are relevant for understanding this phenomenon.\footnote{Another option is that the states only become equivalent as perceived by outside observers, as required by the central dogma \cite{Almheiri:2020cfm}, but perhaps interior observers could still distinguish them?}

\subsection*{Strong coupling}

At strong coupling when $g_{i j}$ become big, the gravitational picture of section \ref{sect:grav} is modified, since in the approximation from \eqref{Weingarten} to \eqref{aproxxx} and \eqref{aprox2}, we assume that terms with $n$ of order $L$ are suppressed. When the coupling become big, that assumption is no longer valid; and so neither is the approximation. This Gaussian approximation fails because for $n\,\,\propto\,\,L$, the contributions from subleading Weingarten functions are not obviously suppressed \cite{drejorrit}. For example, we can no longer trust the scaling formula
\begin{equation}
    {\rm Wg}(\a \b^{-1})\,\,\propto\,\,L^{-\#(\alpha\cdot \beta^{-1})}\,,
\end{equation}
which validated the Gaussian approximation. The combinatoric prefactors may also enhance naively subleading contributions at high order in the coupling constant.

This means that the multi-trace contribution in \eqref{Weingarten} might become relevant at strong coupling. One would obtain multi-trace terms in the brane partition function \eqref{iden1}. These are clearly interpreted in gravity as corresponding with multi-local interaction vertices, we could then represent these by also allowing dotted lines connecting dotted vertices and multiple local interaction vertices; which makes the gravitational expansion more involved. Feynman rules for those dotted diagrams contain further information on $\C$ since these rules depend on multi-trace combinations of $\C$.

It would be interesting to obtain analytic control over these multi-trace deformations by scaling the couplings in certain specific ways.

\subsection*{Fixed Hamiltonians}
Finally, we briefly mention the gravitational interpretation of fixing the random Hamiltonians $H$ to one single Hamiltonian matrix $\H$. This was investigated in \cite{drejorrit} using a deformed matrix integral similar to \eqref{modelbis}, but where $H$ is coupled to an external matrix $\H$ with coupling constant $1/\s^2$. 

Whilst not our focus, the gravitational interpretation of non-random matrix elements \eqref{nonrandombis} involves understanding how $\H$ gets encoded in gravity in addition to $\C$. Therefore we briefly summarize the results of \cite{drejorrit}. The gravity interpretation for the eigenvalues of $\H$ only affects the bulk JT gravity spacetime description, not the behavior of EOW branes discussed throughout this work. This is why our discussion in earlier sections decoupled from fixing $\H$.

For weak coupling $1/\s^2$ one finds a deformation of the JT gravity action which can be interpreted as inserting many local defects \cite{Mertens:2019tcm,Mertens:2020hbs,Turiaci:2020fjj,Witten:2020wvy,Maxfield:2020ale}. These are the analogue of the interaction vertices discussed in this work. The associated coupling constants depend on $\H$, in line with point $2$ of the main conclusion.

When the coupling increases, nonlocal bulk spacetime interactions become important, for precisely the same Weingarten reasons, giving a nonlocal dilaton gravity action. The analogue of terms with $n\,\,\propto\,\, L$ becoming important, is that \emph{macroscopic} operator insertions appear. These tear up the smooth spacetime with large holes \cite{kazakov1990simple}.

For strong coupling we approach the eigenbrane picture \cite{Blommaert:2019wfy,Blommaert:2020seb} with many extra macroscopic holes in spacetime. The boundary conditions on these extra holes \cite{Goel:2020yxl} encode the eigenvalues of $\H$. However, the theory with infinitely many eigenbranes is not under good control, and something far more drastic probably happens. Signs were found \cite{drejorrit,das1990new} of some branched polymer phase of gravity, where smooth spacetime is completely broken. Then the question is what replaces smooth spacetime; what is the true microscopic description of gravity? These works build towards \emph{deriving} a concrete microscopic picture.

What remains is the more illusive gravitational interpretation of the random unitaries $U$. These are irrelevant for observables like partition functions, or the spectral form factor; but crucial for correlation functions \cite{Saad:2019pqd,Blommaert:2020seb,Iliesiu:2021ari,Stanford:2020wkf,Blommaert:2020yeo} and density matrix elements. This is an important open problem \cite{Stanford:2021bhl}.

\subsection*{Acknowledgments}
We happily thank Raphael Bousso, Ven Chandrasekaran, Arvin Shahbazi-Moghaddam and Shunyu Yao for discussions. Special thanks to Jorrit Kruthoff for countless interesting discussions on related topics. AB was supported by a BAEF fellowship, by the SITP and by the ERC-COG Grant NP-QFT No. 864583. MU is supported in part by the NSF Graduate Research Fellowship Program under grant DGE1752814, by the Berkeley Center for Theoretical Physics, by the DOE under award DE-SC0019380 and under the contract DE-AC02-05CH11231, by NSF grant PHY1820912, by the Heising-Simons Foundation, the Simons Foundation, and National Science Foundation Grant No. NSF PHY-1748958.

\appendix
\section{Gravitational amplitudes}\label{app:main}
Here we gather some gravitational details relevant in the main text.

\subsection{Pinwheels}\label{app:gravamp}
First consider the pinwheel geometry $Z_n(\beta)$ of \cite{Penington:2019kki}, which is a disk geometry with $n$ pieces of asymptotic boundary of length $\beta$, separated by $n$ boundary segments that describe the geodesic trajectory of some particle of mass $\mu$; these represent the EOW branes. For example
\begin{equation}
    Z_3(\beta)=\quad \raisebox{-13.5mm}{\includegraphics[width=30mm]{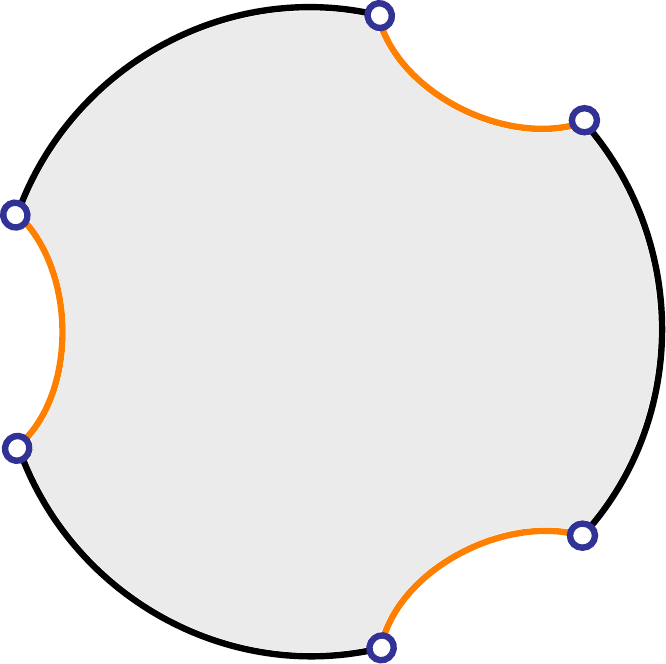}}\quad.
\end{equation}
This amplitude was computed using the techniques of  \cite{Yang:2018gdb,Saad:2019pqd} and gives
\begin{equation}
    Z_n(\beta)=e^{\S}\int_0^{+\infty} \d E\,\exp(-n\beta E )\,F(E)^n\,\frac{1}{4\pi^2}\sinh(2\pi E^{1/2})\quad,\quad F(E)= \Gamma\left(\mu-1/2\pm \i E^{1/2}\right)\label{pinwheel}
\end{equation}
We note in passing that this formula is also easily derivable in the BF formalism of \cite{Blommaert:2018iqz,Blommaert:2018oro}, where the mass $\mu$ boundary particles are represented by Wilson lines and one recognizes $F(E)$ as the $3j$ symbols with one trivial representation; because there is nothing on the other side of the particle.

In the matrix integral this pinwheel corresponds with the observable
\begin{equation}
    \Tr(e^{-\beta H}F(H)\dots e^{-\beta H}F(H))=\int_{-\infty}^{+\infty} \d E\,\exp(-n\beta E )\,F(E)^n\,\Tr \delta(E-H)\,.
\end{equation}
The leading order expectation value of $\Tr \delta(E-H)$ equals the disk amplitude with fixed energy boundary conditions \cite{Saad:2019lba,Blommaert:2019wfy}
\begin{equation}
    \average{\Tr \delta(E-H)}=\frac{e^{\S}}{4\pi^2}\sinh(2\pi E^{1/2})=\rho(E)\,,
\end{equation}
which indeed reproduces \eqref{pinwheel}. Including handles on the pinwheel replaces the genus zero disk answer \eqref{pinwheel} with
\begin{equation}
    \int_{-\infty}^{+\infty} \d E\,\exp(-n\beta E )\,F(E)^n\,\average{\rho(E)}\,,
\end{equation}
where $\average{\rho(E)}$ is the exact spectral density in the matrix integral. This can be calculate order per order in the genus expansion using Weil-Peterson volumes, and nonperturbatively using D-branes \cite{Saad:2019lba,Blommaert:2019wfy}. 

When there are two pinwheels, we must include spacetime wormholes that connect them. Summing over all genus gives rise to the full spectral correlation $\average{\rho(E_1)\rho(E_2)}$ of random matrix theory \cite{Saad:2019lba,Blommaert:2019wfy,mehta2004random}
\begin{equation}
    \average{Z_{n_1}(\beta)Z_{n_2}(\beta)}= \int_{-\infty}^{+\infty} \d E_1\,\exp(-n_1 \beta E_1 )\,F(E_1)^{n_1}\, \int_{-\infty}^{+\infty} \d E_2\,\exp(-n_2 \beta E_2 )\,F(E_2)^{n_2}\,\average{\rho(E_1)\rho(E_2)}\,,
\end{equation}
which, for all intents and purposes, can be approximated as
\begin{equation}
    \average{\rho(E_1)\rho(E_2)}=\rho(E_1)\rho(E_2)+\delta(E_1-E_2)\rho(E_1)-\frac{\sin(\pi \rho(E)(E_1-E_2))^2}{\pi^2(E_1-E_2)^2}\,.\label{speccor}
\end{equation}
The generalization to multiple pinwheels is obvious \cite{Blommaert:2019wfy}. 

Though these expressions are very explicit, it is more practical to work in a microcanonical ensemble, where things simplify even further. In some microcanonical ensemble centered around $\E$, one computes for example\footnote{We have in mind some implicit Gaussian weight centered around $\E$ which defines the energy bin smoothly.}
\begin{equation}
    \average{Z_{n_1}(\E)Z_{n_2}(\E)}= \int_{\E} \d E_1\,F(E_1)^{n_1}\, \int_{\E} \d E_2\,F(E_2)^{n_2}\,\average{\rho(E_1)\rho(E_2)}=F(\E)^{n_1+n_2}\,e^{2\Ss}\,,\label{pinwheelmicro}
\end{equation}
where the total number of eigenvalues in this microcanonical bin computes the microcanonical entropy
\begin{equation}
    \int_{\E} d E_1\,\rho(E_1)=e^{\Ss}\,.
\end{equation}
The second equality in \eqref{pinwheelmicro} follows from the definition of the microcanonical ensemble, the width of the energy bin is much smaller than $1$ but much bigger than the typical level spacing $1/\rho(E)$. The function $F(E)$ varies on energy scales of order $1$ and can therefore be approximated as constant within the bin. Furthermore, on energy scales bigger than $1/\rho(E)$ the sine kernel in \eqref{speccor} is essentially indistinguishable from the Dirac delta term; and these two contributions therefore cancel out, and to good approximation
\begin{equation}
    \int_{\E} \d E_1\int_{\E} \d E_2\,\average{\rho(E_1)\rho(E_2)}=\int_{\E} \d E_1\,\rho(E_1)\int_{\E} \d E_1\,\rho(E_1)=e^{2\Ss}
\end{equation}

Using these results one computes, with the rules explained in section \ref{sect:review}
\begin{equation}
    \rho_{i j}=\delta_{i j}\,F({\E})\,e^{\Ss}\quad,\quad
    \rho_{i j}\,\rho_{k l}=\delta_{i j}\,\delta_{k l}\,F({\E})^2\,e^{2\Ss}+\delta_{i l}\,\delta_{k j}\,F({\E})^2\,e^{\Ss}\,.
\end{equation}
These are the results mentioned in \eqref{rho} and \eqref{rhorho}. Notice that the details of the EOW brane boundary conditions, captured by the kernel $F({\E})$, are just overall normalization constants in these amplitudes; all this dependence drops out when we consider normalized density matrices, and compute normalizes quantities like Renyi entropies. Life in the microcanonical ensemble is simple.

\subsection{Modeling interactions}\label{app:marking}
We gather formulas about disk amplitudes with marking operators in minimal strings and in JT gravity, more details are contained in \cite{Mertens:2020hbs,Mertens:2020pfe,Hosomichi:2008th,Kostov:2002uq}. 

First consider a circular FZZT boundary \cite{Saad:2019lba,Maldacena:2004sn,Fateev:2000ik,Ponsot:2001ng}, without marked points, which corresponds in random matrix theory with
\begin{equation}
    \Tr(\log(E-H))=\quad \raisebox{-5.5mm}{\includegraphics[width=12.5mm]{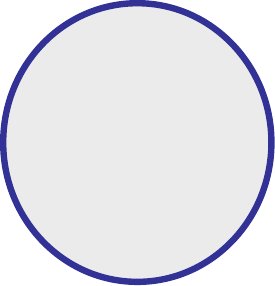}}\quad\,.
\end{equation}
The random matrix observable includes a sum over genus in gravity, which we suppress for presentation purposes. Now consider the minimal string boundary three point function of marking operators\footnote{Boundary chiral vertex operators have three labels \cite{moore1989classical}, the two extra labels denote the boundary states between which they intertwine. In string language these are the Chan-Patton indices for the two D-branes between which the open string operator stretches, generalized to non-coincident D-branes.}
\begin{equation}
    \average{\mathcal{T}_{1\,E_1E_2}\mathcal{T}_{1\,E_2E_3}\mathcal{T}_{1\,E_3E_1}}\,.
\end{equation}
As written the marking operators intertwine between segments with FZZT boundary states respectively $E_1$, $E_2$ and $E_3$.\footnote{FZZT boundary conditions are technically a double cover of the energy axis and should be labeled by $z$ with $E=-z^2$, there is a unique Liouville primary corresponding with each $z$; this is not relevant here so we suppress it for reader comfort.} This corresponds in random matrix theory with the following observable \cite{Mertens:2020hbs,Mertens:2020pfe,Hosomichi:2008th,Kostov:2002uq}
\begin{equation}
    \Tr(\frac{1}{E_1-H}\frac{1}{E_2-H}\frac{1}{E_3-H})=\quad \raisebox{-7.5mm}{\includegraphics[width=18mm]{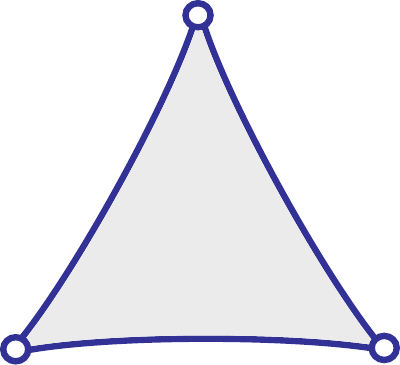}}\quad .\label{markings}
\end{equation}

Now consider a disk with thermal boundary length $\alpha_1+\beta_1+\alpha_2+\beta_2+\alpha_3+\beta_3$, which corresponds in random matrix language with  $\Tr(\exp(-(\alpha_1+\beta_1+\alpha_2+\beta_2+\alpha_3+\beta_3)H))$. Laplace transforming some segment of thermal boundary gives a segment of FZZT boundary, this is obviously true from the matrix integral formulas. Therefore we have the correspondence
\begin{equation}
    \Tr(\frac{1}{E_1-H}e^{-\beta_1 H}\frac{1}{E_2-H}e^{-\beta_2 H}\frac{1}{E_3-H}e^{-\beta_3 H})=\quad \raisebox{-13.5mm}{\includegraphics[width=30mm]{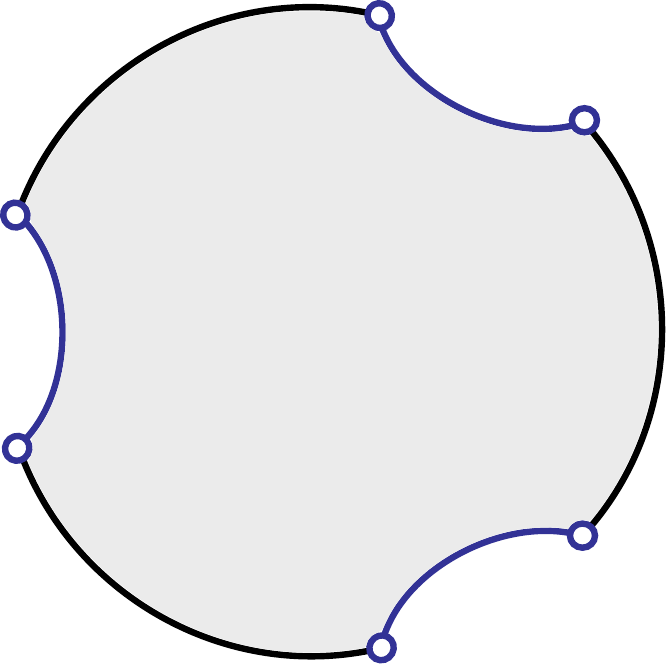}}\quad\,.\label{pin}
\end{equation}
The gravity amplitude mimics the pinwheel amplitude considered in appendix \ref{app:gravamp} and in \cite{Penington:2019kki}, but with FZZT boundary conditions instead of mass $\mu$ boundary conditions. FZZT and mass $\mu$ boundary states are linear combination of each other; one checks that their JT boundary wave functions form complete sets for certain complex ranges of $E$ respectively $\mu$ \cite{Saad:2019pqd,Penington:2019kki}, hence there is a basis transform between them.

Now we see that taking the thermal length of one of the segments in the pinwheel to zero reproduces amplitudes of the type \eqref{markings}. The random matrix dual clarifies that this is indeed the correct limit to take, if you send $\beta_1$, $\beta_2$ and $\beta_3$ to zero in \eqref{pin}, you recover \eqref{markings}. This proves that the dilaton gravity interpretation of a marking operator $\mathcal{T}_{1\,E_1E_2}$ corresponds with a piece of thermal boundary sandwiched between FZZT boundary segments with boundary conditions $E_1$ and $E_2$, where the thermal length of the sandwiched segments is taken to zero.

The generalization to mass $\mu$ boundaries is straightforward, since these are just linear combinations of FZZT boundaries. We can consider for example the minimal string boundary three point function\footnote{In stringy language the $k$ flavors of interior modes are ordinary Chan-Paton indices, because all D-branes coincide at $\mu$.}
\begin{equation}
    \average{\mathcal{T}_{1\,\mu\mu}\mathcal{T}_{1\,\mu\mu}\mathcal{T}_{1\,\mu\mu}}\,.
\end{equation}
Following the above, this corresponds in random matrix theory with the observable
\begin{equation}
    \Tr\Big(\Gamma\left(\mu-1/2\pm \i H^{1/2}\right)\Gamma\left(\mu-1/2\pm \i H^{1/2}\right)\Gamma\left(\mu-1/2\pm \i H^{1/2}\right)\Big)=\quad \raisebox{-7.5mm}{\includegraphics[width=18mm]{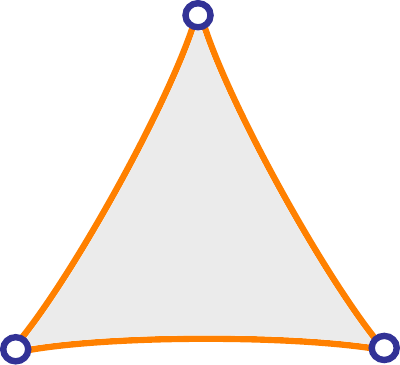}}\quad\,,
\end{equation}
and therefore in gravity with the $\beta=0$ limit of the pinwheel diagrams studied in appendix \ref{app:gravamp}.

Notice that taking the $\beta=0$ limit of the pinwheels, gives a finite answer for the second diagram in \eqref{dbranepart1}; representing a disk ending on a mass $\mu$ particle with a single marking operator inserted. Naively one might have thought that amplitude would vanish, since the boundary is a geodesic, but it does not.

In summary, if we model interactions by insertions of marking operators, we know the corresponding observables in random matrix theory and the corresponding boundary conditions in gravity, and we can compute all amplitudes we want. One could consider modeling interactions by more general minimal open string Tachyons $\mathcal{T}_{n\,\mu\mu}$, however the random matrix dual of the corresponding boundary correlators is not actually known,\footnote{Recently an educated guess was made in \cite{Mertens:2020pfe} which perhaps deserves further study.} and concordantly neither are the precise boundary conditions in dilaton gravity. We expect the conclusions of this work to hold when working with these other models for interactions.

\subsection{Computing amplitudes}\label{app:varia}
Here we go through the JT gravity calculation for one amplitude that contributes to the matrix element. The example which we choose is sufficiently complex so that the generalization to all amplitudes should be straightforward. We consider \eqref{hole}
\begin{equation}
    \bra{\psi_j}\ket{\psi_i}\supset \frac{1}{3}\, G^4\,\Big(g^3\Big)_{i j}\,G^3\,\Tr\Big(g^3\Big)\quad \raisebox{-10mm}{\includegraphics[width=18mm]{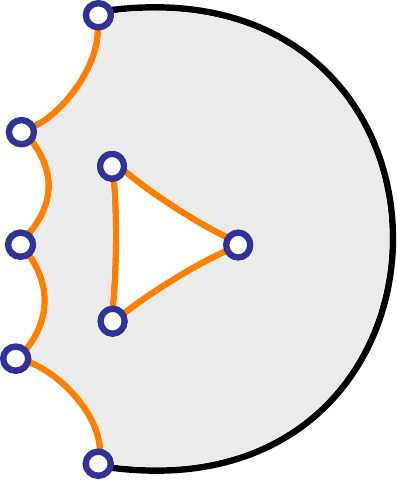}}\quad\,,
\end{equation}
where we already extracted the EOW particle Feynman rules, such that the diagram reflects a pure JT gravity calculation. The way to proceed is to first treat each boundary loop as analogous to a standard fixed length boundary; chopping up the surface by cutting off ``trumpets'' ending on each boundary \cite{Saad:2019lba}, and on the unique geodesic inside the Riemann surface homologous to the boundary in question. 

The remaining amputated amplitude with geodesic boundaries computes the Weil-Petersson volume, which can be calculated by further chopping up this Riemann surface into three holed spheres. This is explained in great detail in \cite{Stanford:2019vob,Saad:2019pqd,Blommaert:2020seb,Iliesiu:2021ari,mirzakhani2007simple,Dijkgraaf:2018vnm,Saad:2019lba}, and will not be repeated here. The newer ingredient is computing the trumpet ending on a boundary circle that includes interacting EOW branes.

Let us work through this in the above example. Cutting the Riemann surface on the blue geodesic of length $b$ leaves two trumpets, here there remains no amputated surface and we need no Weil-Petersson volumes
\begin{equation}
    \quad \raisebox{-10mm}{\includegraphics[width=18mm]{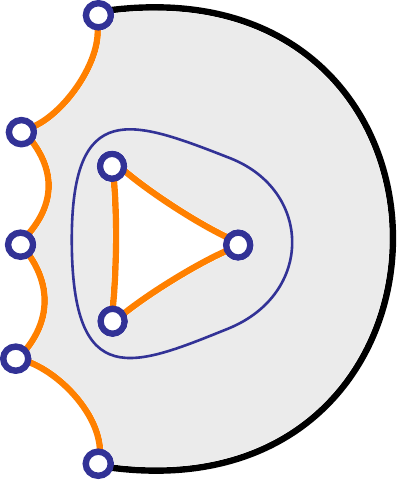}}\quad=\int_0^\infty \d b\,b
    \quad \raisebox{-10mm}{\includegraphics[width=32mm]{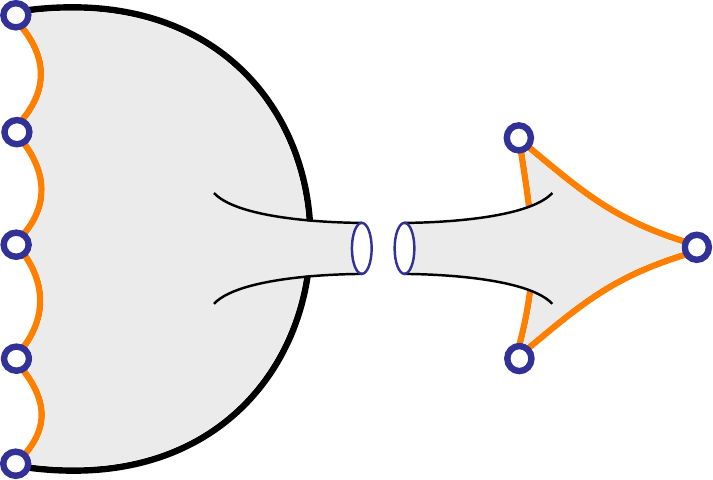}}\quad\,.
\end{equation}
Each of the remaining pieces is a trumpet with one geodesic boundary and one boundary that involves segments of EOW particles.

As explained in appendix \ref{app:gravamp} and appendix \ref{app:marking}, without geodesic boundary, the amplitudes could be easily calculated using the boundary particle formalism \cite{Yang:2018gdb,Saad:2019pqd} or the BF formulation \cite{Blommaert:2018iqz,Blommaert:2018oro,Mertens:2019tcm,Mertens:2018fds,Iliesiu:2019xuh}
\begin{equation}
    \quad \raisebox{-10mm}{\includegraphics[width=14mm]{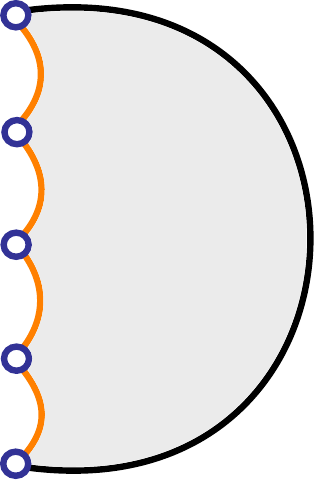}}\quad=\int_{0}^{+\infty}\d E\,\exp(-\beta E)\,F(E)^4\,\frac{e^{\S}}{4\pi^2}\sinh(2\pi E^{1/2})\,.
\end{equation}
Including the geodesic boundary is easy within the BF or first order formulation, where it is interpreted as introducing a hyperbolic defect; amplitude wise this simply replaces the sinh factor with a cosine \cite{Mertens:2019tcm}
\begin{equation}
    \quad \raisebox{-10mm}{\includegraphics[width=17mm]{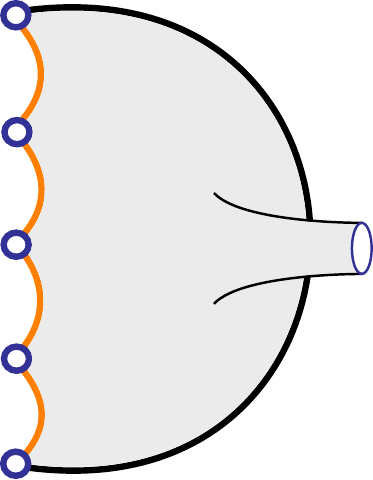}}\quad=\int_{0}^{+\infty}\d E\,\exp(-\beta E)\,F(E)^4\,\frac{1}{2\pi}\frac{1}{E^{1/2}}\cos(b E^{1/2})\,.
\end{equation}
The trumpet with one geodesic boundary and one triangle boundary therefore becomes
\begin{equation}
    \quad \raisebox{-5.4mm}{\includegraphics[width=14.5mm]{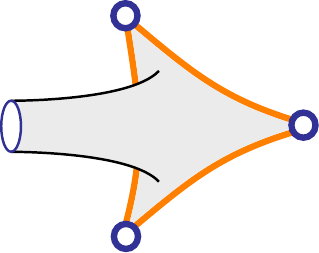}}\quad=\int_{0}^{+\infty}\d E\,F(E)^3\,\frac{1}{2\pi}\frac{1}{E^{1/2}}\cos(b E^{1/2})\,.
\end{equation}
We can now immediately compute the $b$ integral using
\begin{equation}
    \int_0^\infty \d b\,b\,\frac{1}{2\pi}\frac{1}{E_1^{1/2}}\cos(b E_1^{1/2})\,\frac{1}{2\pi}\frac{1}{E_2^{1/2}}\cos(b E_2^{1/2})=-\frac{1}{4\pi^2}\frac{E_1+E_2}{E_1^{1/2}E_2^{1/2}}\frac{1}{(E_1-E_2)^2}=\average{\rho(E_1)\rho(E_2)}_0\,,
\end{equation}
which is the genus zero contribution to the spectral correlation \cite{Saad:2019lba}. Combining the elements, we obtain
\begin{equation}
    \quad \raisebox{-10mm}{\includegraphics[width=18mm]{224_40.5.pdf}}\quad=\int_{0}^{+\infty}\d E_1\,\exp(-\beta E_1)\,F(E_1)^4\,\int_{0}^{+\infty}\d E_2\,F(E_2)^3\,\average{\rho(E_1)\rho(E_2)}_0\,.
\end{equation}
Including any number of handles and nonperturbative effects in that genus expansion simply replaces the genus zero connected spectral correlator with the full correlator \eqref{speccor} of random matrix theory \cite{Blommaert:2020seb,Saad:2019pqd,Iliesiu:2021ari}. The generalization to arbitrary amplitudes should now be obvious.

Another new application of the BF formulation is the calculation of \cite{Gao:2021uro}, which considers a trumpet with mass $\mu$ particle on the geodesic boundary. In the BF formulation, massive particles become Wilson lines, and the mass $\mu$ labels discrete series irreducible representations of SL$(2,$R$)$. The geodesic length $b$, over which is integrated, labels hyperbolic conjugacy class elements of SL$(2,$R$)$, and Wilson lines in the conjugacy class element basis contribute characters to BF amplitudes \cite{Mertens:2018fds,Blommaert:2018iqz,Blommaert:2018oro,witten1991quantum}. One then finds
\begin{equation}
    \quad \raisebox{-10mm}{\includegraphics[width=22.5mm]{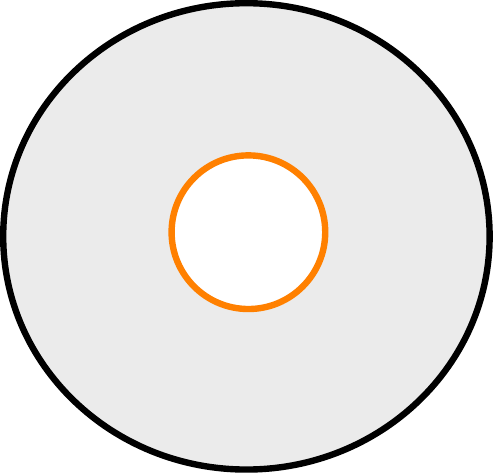}}\quad = \int_0^\infty \d b\,\chi_\mu(b) \int_{0}^{+\infty}\d E\,\exp(-\beta E)\,\frac{1}{2\pi}\frac{1}{E^{1/2}}\cos(b E^{1/2})\,,
\end{equation}
and the discrete series characters evaluated on hyperbolic conjugacy class elements are in this convention \cite{vilenkin1978special}
\begin{equation}
    \chi_\mu(b)=\frac{\exp(-\mu b)}{2\sinh(b/2)}\,.
\end{equation}
This therefore indeed reproduces formula (2.47) of \cite{Gao:2021uro}, there obtained via direct canonical quantization. This character formula is also relevant when exactly computing the contributions of matter loops around handles in JT gravity. Inserting it as an extra kernel in the double trumpet gives the annulus with one matter loop going around. More loops are annoying since the particles can then cross, giving potential SL$(2,$R$)$ $6j$ symbols \cite{Blommaert:2018iqz,Blommaert:2018oro,Iliesiu:2019xuh,Blommaert:2018oue}. Could these be used to study deviations from random matrices \cite{Stanford:2021bhl}?

\bibliographystyle{ourbst}
\bibliography{Refs}
\end{document}